\documentclass[a4paper,prd,preprintnumbers,twocolumn,superscriptaddress,nofootinbib,amsmath,amssymb]{revtex4-2}

\usepackage{graphicx} % Required for inserting images

\usepackage{hyperref}
\hypersetup{
  colorlinks=true,        % false: boxed links; true: colored links
  linkcolor=blue,         % color of internal links
  citecolor=cyan,         % color of links to bibliography
}

\usepackage{graphicx}% Include figure files
\usepackage{dcolumn}% Align table columns on decimal point
\usepackage{bm}% bold math
\usepackage{amsmath}
\usepackage{amssymb}

\newcommand{\beq}{\begin{equation}}
\newcommand{\eeq}{\end{equation}}
\newcommand{\bea}{\begin{eqnarray}}
\newcommand{\eea}{\end{eqnarray}}

\usepackage{bbm}
\usepackage{amsfonts}
\usepackage{mathrsfs}
\usepackage{latexsym}
\usepackage{epsfig}
\usepackage{epstopdf}
\usepackage{epstopdf}
\usepackage{graphicx}
\usepackage{subcaption}
\usepackage{amssymb}
\usepackage{amsmath}
\usepackage{ragged2e}
\usepackage{dcolumn}
\usepackage{bm}
\usepackage{comment}
\usepackage{xcolor}

\begin{document}

%\title{Klein–Gordon Perturbations and Bound-State Spectra of massive vector and scalar fields in a black hole spacetime with a topological defect}

%\title{Quasinormal mode spectra and bound-state configurations of massive vector and scalar fields in a black hole spacetime with a topological defect}

%\title{Massive Vector and Scalar Fields around a Black Hole with a Topological Defect: Quasinormal Modes and Bound States in a Plasma Medium}

\title{Scalar and Electromagnetic Perturbations around a Black Hole with a Topological Defect: Quasinormal Modes and Quasi-bound States in a Plasma Medium}

\author{Dilmurod Umarov}
\email{dilmurodumarov666@gmail.com}
\affiliation{Institute of Fundamental and Applied Research, National Research University TIIAME, Kori Niyoziy 39, Tashkent 100000, Uzbekistan} 

\author{Farruh~Atamurotov}
\email{atamurotov@yahoo.com}
\affiliation{Kimyo International University in Tashkent, Shota Rustaveli str. 156, Tashkent 100121, Uzbekistan}
\affiliation{Research Center of Astrophysics and Cosmology, Khazar University, 41 Mehseti Street, Baku AZ1096, Azerbaijan}

\author{Ahmadjon~Abdujabbarov}
\email{ahmadjon@astrin.uz}
\affiliation{School of Physics, Harbin Institute of Technology, Harbin 150001, People’s Republic of China}
\affiliation{Institute of Fundamental and Applied Research, National Research University TIIAME, Kori Niyoziy 39, Tashkent 100000, Uzbekistan} 
\affiliation{University of Tashkent for Applied Sciences, Str. Gavhar 1, Tashkent 100149, Uzbekistan}

\author{Chengxun Yuan}
   \email{yuancx@hit.edu.cn}
\affiliation{School of Physics, Harbin Institute of Technology, Harbin 150001, People’s Republic of China}

\author{G.~Mustafa}
\email{gmustafa3828@gmail.com}
\affiliation{Department of Physics, Zhejiang Normal University, Jinhua 21004, China}

\begin{abstract}
We investigated the influence of a plasma environment on the optical and perturbative properties of a black hole with a topological defect, characterized by the parameter \(k\). We first established a straightforward correspondence between the real part of the quasinormal-mode (QNM) frequencies in the eikonal limit and the black-hole shadow radius. We then demonstrated that the Lyapunov exponent associated with the photon sphere exhibits only a weak dependence on the plasma frequency, while it monotonically decreases as the topological-defect parameter \(k\) increases. Subsequently, we analyzed massive scalar-field perturbations by deriving the associated effective potential and computing the QNM spectrum using the third- and sixth-order WKB approximations for both homogeneous and radially inhomogeneous plasma configurations, including the singular isothermal sphere (SIS) and non-singular isothermal sphere (NSIS) density profiles. Our results show that the presence of plasma induces shifts in both the oscillation frequencies and the damping rates of the modes, and that larger values of \(k\) systematically suppress the real part of the QNM frequencies. Among the plasma models considered, the NSIS profile generally yields slightly higher oscillation frequencies than both the SIS and homogeneous cases. Finally, we derived the dynamical equations governing electromagnetic perturbations in a cold, unmagnetized plasma and demonstrated that the axial and polar sectors decouple. In the axial sector, the plasma frequency enters as an effective mass term, thereby permitting the existence of quasi-bound states only in the case of a homogeneous plasma and only when the plasma frequency lies below a critical threshold that depends on the topological-defect parameter \(k\) and the multipole index \(l\).
\end{abstract}

\maketitle

\section{Introduction}

The advent of gravitational wave astronomy and direct black hole imaging has fundamentally transformed our capacity to investigate the strong-field regime of gravity. In particular, the detections of gravitational waves from compact binary coalescences by the LIGO and Virgo collaborations~\cite{Abbott:2016blz,Abbott:2017vtc}, together with the horizon scale imaging of the supermassive black hole shadows M87* and Sgr A* by the Event Horizon Telescope (EHT)~\cite{Akiyama:2019eap,Akiyama:2022cmc}, have opened unprecedented observational windows onto the Universe. These observations not only provide stringent tests of general relativity, but also serve as powerful laboratories for investigating the nature of dark matter, the properties of plasma environments, and potential signatures of physics beyond the Standard Model~\cite{Psaltis:2019hip,Vagnozzi:2022moj}.

Among the nontrivial configurations anticipated within the frameworks of high-energy particle physics and early Universe cosmology, topological defects constitute a particularly significant class of theoretical structures. These defects, such as cosmic strings, domain walls, and monopoles, are expected to form during symmetry-breaking phase transitions~\cite{Kibble:1976sj,Vilenkin:1984ib}. Global monopoles, arising from the spontaneous rupture of a global $O(3)$ symmetry, represent a particularly interesting class of such defects. Their gravitational field, when coupled to a black hole, yields a static, spherically symmetric solution first studied by  Barriola and Vilenkin~\cite{Barriola:1989hx}. This solution is characterized by a dimensionless parameter $k$ in the metric function, which is directly proportional to the square of the symmetry-breaking scale $\eta$, i.e. $k = 8\pi \eta^2$. Recent analyzes of EHT observations have placed stringent upper bounds on this parameter $k \lesssim 0.005$ (1$\sigma$), thus constraining the energy scale of the underlying phase transition~\cite{Vagnozzi:2022moj}.

Astrophysical black holes in realistic settings are not isolated systems embedded in vacuum. Instead, they are typically situated within complex, high density environments such as accretion disks, stellar winds, and dark matter halos that frequently contain a plasma component, which may be either magnetized or unmagnetized. The presence of a plasma significantly alters the propagation of light and other fields, leading to observable modifications in both the black hole shadow and its perturbation spectrum. For instance, the refractive index of the plasma modifies the effective geometry for photon trajectories, affecting the size and shape of the black hole shadow~\cite{Bisnovatyi:2010a,Perlick:2015vta,Perlick:2017cjo,Atamurotov:2015nra,Babar:2020txt,Atamurotov:2021cgh,Atamurotov:2022iwj,Atamurotov:2022nim}. Furthermore, the characteristic QNMs that govern the relaxation of a black hole after a perturbation are also sensitive to the surrounding matter~\cite{Konoplya:2011qq,Konoplya:2021ube,Konoplya:2022hbl}.

The interaction between black hole perturbations and surrounding plasma environments has attracted increasing attention in recent years. In the eikonal limit, corresponding to large values of the multipole number, a well-established relation links the QNM frequencies to the characteristics of unstable null geodesics at the photon sphere: the real part of the QNM frequency is proportional to the angular velocity of the corresponding circular null orbit, while the imaginary part is proportional to its Lyapunov exponent~\cite{Cardoso:2008bp,1984PhRvD..30..295F}. This correspondence has been further developed to establish a direct relation between the real part of the eikonal QNM frequencies and the radius of the black hole shadow~\cite{Jusufi:2019ltj,Jusufi:2020wmp,Stefanov:2010xz,Cuadros-Melgar:2020,Yang:2021}, and has also been extended to rotating spacetimes~\cite{Li:2021ptl,Wu:2022xkj}. Analytic expressions for the QNM spectrum in the eikonal regime have likewise been derived~\cite{Churilova:2019ehh}. It is important to emphasize, however, that this correspondence is not universally applicable and can be violated in certain modified theories of gravity, such as Einstein Lovelock gravity~\cite{Konoplya:2017wot}. Moreover, the presence of a dispersive plasma medium can alter the QNM shadow relation through its effect on the refractive index, and can also lead to additional phenomena. In particular, the effective mass associated with the plasma frequency, combined with the gravitational potential well, can support quasi-bound states of electromagnetic waves~\cite{Rosa:2011my,Cannizzaro:2020uap}.

%The interplay between black hole perturbations and plasma environments has been a subject of growing interest. In the eikonal limit, corresponding to large multipole numbers, a well-established relation links the QNM frequencies to the parameters of unstable null geodesics at the photon sphere: the real part is proportional to the angular velocity, and the imaginary part to the Lyapunov exponent~\cite{Cardoso:2008bp}. This connection has been extended to include the effects of a plasma medium, providing a direct link between QNMs and the shadow radius~\cite{Jusufi:2019ltj,Stefanov:2010xz}. Moreover, the presence of a plasma can give rise to new phenomena, such as the existence of quasi-bound states for electromagnetic waves, which form due to the effective mass induced by the plasma frequency and the gravitational potential well~\cite{Rosa:2011my,Oliveira:2023xjk}.

Topological defects and plasma have each been studied for their impact on black hole observables, but a unified, in-depth treatment of black hole spacetimes that actually contain topological defects remains largely missing. This work closes that gap with a systematic exploration of how a surrounding plasma reshapes both the observable signatures and the perturbation spectra of a black hole threaded by a topological defect. We focus on four key questions: (1) How do different plasma configurations homogeneous and inhomogeneous distort the established link between the black hole shadow radius and the eikonal QNM spectrum? (2) How do the topological defect parameter \(k\) and the presence of plasma jointly modify the instability timescale of null geodesics, encoded in the Lyapunov exponent? (3) In what precise and quantitative ways do concrete plasma models shift the homogeneous SIS and the NSIS scalar QNM frequencies? (4) Can the plasma in this spacetime trap electromagnetic waves into quasi-bound states, and if so, how do their characteristic frequencies depend on \(k\) and on the plasma density profile?

To address these challenges in a systematic manner, we employ a multi-faceted methodological framework. First, we analyze the photon sphere and its associated Lyapunov exponent in order to quantitatively establish the relationship between QNMs and the shadow radius in a homogeneous plasma environment. Next, we carry out a high-precision study of scalar-field perturbations, utilizing the Wentzel–Kramers–Brillouin (WKB) approximation extended to sixth order and incorporating homogeneous, SIS, and NSIS plasma configurations. Finally, we derive the full set of dynamical equations governing electromagnetic wave propagation in a cold, unmagnetized plasma on this background, perform a decoupling of axial and polar perturbative sectors, and determine the exact conditions under which quasi-bound states can form in the axial sector.

This paper is structured as follows. In Sect.~\ref{sec:bh_metric}, we review the metric of a black hole with a topological defect. Sec.~\ref{sec:shadow_qnm} is dedicated to exploring the connection between the shadow radius and QNMs in the presence of a plasma, including an analysis of the Lyapunov exponent. In Sect.~\ref{sec:scalar_qnm}, we investigate the QNMs of a massive scalar field, considering homogeneous and inhomogeneous plasma profiles (SIS, NSIS). Sect.~\ref{sec:em_plasma} derives the fundamental equations for photon-plasma interactions in curved spacetime, decouples the axial and polar sectors, and analyzes the effective potentials and quasi-bound states for electromagnetic perturbations. Finally, we summarize our findings and discuss their observational implications in Sect.~\ref{sec:conclusion}. Some detailed analytical computations have been delegated to Appendix \ref{Ap1}, Appendix \ref{Ap2} and Appendix \ref{AppC}, respectively.

\textit{Notations and conventions:} Throughout the paper, we will use units in which \( G = 1 = c \) and the signature of the flat spacetime Minkowski metric will be taken to be mostly positive, i.e., $\eta_{\mu\nu} = \operatorname{diag}(-1,1,1,1).$ Moreover, all the Greek indices \( \mu,\nu,\alpha,\dots \) denote spacetime indices, and we will exclusively work with four spacetime dimensions.

\section{ Black hole with a topological defect \label{sec:bh_metric}}

Topological defects play an important role in several different contexts of physical interest and can likewise be produced in a variety of scenarios, including during phase transitions in the early Universe \cite{Hindmarsh:1993av} or as a result of the spontaneous breaking of symmetries \cite{Vilenkin:1981zs,Vilenkin:1981kz,Vilenkin:1982hm,Vilenkin:1984ib}. For instance, global monopoles may arise from the breaking of a global $O(3)$ symmetry down to a $U(1)$ subgroup, which can be realized, for example, by a triplet of scalar fields. The stability of such configurations has been investigated in a number of works \cite{Goldhaber:1989na,Rhie:1990kc,Perivolaropoulos:1991du}. The gravitational field of a Schwarzschild black hole carrying a global monopole charge was first studied in Refs. \cite{Barriola:1989hx,Dadhich:1997mh}, and the corresponding black hole solution is described by the following metric function:
\begin{align}
    f(r)=1-k-\frac{2M}{r}\,,\label{1a}
\end{align}
where the precise physical interpretation of the parameter $k$ is model dependent. As an illustrative example, one may consider a triplet of real scalar fields $\phi^i$ transforming under a global $O(3)$ symmetry, whose dynamics are governed by the following Lagrangian:
\begin{align}
    \mathcal{L}=\frac{1}{2}(\partial\phi^i)^2-\frac{\mathcal{\lambda}}{4}(\phi^i\phi^i-\eta^2)^2\,,
\end{align}
for which the associated global monopole is given by:
\begin{align}
\phi^i = \eta f(r) \frac{x^a}{|\overline{\mathbf{x}}|}\,,
\end{align}
with $f(r)\rightarrow1$ as $|\overline{\mathbf{x}}|\rightarrow0$, and the parameter $k$ in Eq. (\ref{1a}) specified by $k=8\pi \eta^2$. In this framework, $k$ is directly associated with the amplitude of the monopole field. The Event Horizon Telescope (EHT) observations impose constraints of $k \lesssim 0.005$ ($1\sigma$) and $k\lesssim 0.1$ ($2\sigma$), which can be recast as upper bounds on the amplitude of the global monopole field, namely $\eta \lesssim 0.014$ and $\eta \lesssim 0.06$, respectively \cite{Vagnozzi:2022moj}.

\section{Connection Between the Shadow Radius and QNMs in the Presence of a Plasma Medium \label{sec:shadow_qnm}}

QNMs represented the characteristic oscillations of the system that carried important information about the stability of the black hole under small perturbations. To analyze them, it was necessary to impose appropriate boundary conditions only an outgoing wave was considered at infinity, while only an ingoing wave was allowed at the horizon. In general, QNMs were described by a complex frequency, where the real part determined the oscillation frequency and the imaginary part accounted for the damping of the modes \cite{Konoplya:2017wot, Jusufi:2019ltj}
\begin{align}
    \omega_{\text{QNM}}=\omega_{\mathcal{\Re}}-i\omega_\mathcal{\Im}\,.
\end{align}

In the eikonal regime, the QNMs of black holes are intimately linked to the properties of unstable null geodesics and the black hole shadow. As established by \cite{Cardoso:2008bp,1984PhRvD..30..295F}, the real part of the QNMs frequencies corresponds to the angular velocity $\Omega_c$ of the last circular photon orbit, while the imaginary part is determined by the Lyapunov exponent $\lambda$, governing the instability timescale of the orbit. This relationship is expressed as 
\begin{equation}
\omega_{\text{QNM}} = \Omega_c l - i\left(n + \frac{1}{2}\right)|\lambda|\,,\label{1}
\end{equation}
and holds not only for static, spherically symmetric spacetimes but also for equatorial orbits in rotating black hole geometries. Further investigations by \cite{Stefanov:2010xz} revealed a connection between eikonal QNMs and gravitational lensing in the strong deflection limit, where $\Omega_c$ and $\lambda$ can be expressed in terms of the observer-lens distance $D_{\text{OL}}$ and the angular position $\theta$ of the lensed image \cite{Jusufi:2019ltj}.  We can write these relations in the presence of a plasma medium as follows [here we shall introduce temporary the speed of light $c$]
\begin{align}
    \Omega_c=\frac{c}{\theta D_{OL}}\sqrt{1-\frac{\omega_{\rm pl}^2(r)}{\omega_{\rm ph}^2(r)}}, \quad \lambda=\frac{c\text{ln}\tilde{r}}{2\pi\theta D_{OL}}\sqrt{1-\frac{\omega_{\rm pl}^2(r)}{\omega_{\rm ph}^2(r)}}\,, \label{2}
\end{align}
where, $\omega_{\rm pl}$ and $\omega_{\rm ph}$ are the plasma frequency and the photon frequency, respectively. However, this correspondence is strictly valid only for high multipole numbers $l$ and certain test fields, as demonstrated by \cite{Konoplya:2017wot}.

The angular radius of the black hole shadow, as measured by an observer located in the asymptotically distant region, can be expressed in terms of the shadow radius and the observer’s distance to the black hole \cite{Perlick:2021aok} as:
\begin{equation}
    \theta=\frac{R_S}{D}\,.\label{3}
\end{equation}
Now, using relations (\ref{1}), (\ref{2}), and (\ref{3}), we find that in the eikonal regime, the real part of QNMs is inversely proportional to the shadow radius and directly proportional to the plasma refractive index
\begin{equation}
\omega_{\mathcal{\Re}} \propto \frac{1}{R_S} 
\sqrt{1-\frac{\omega_{\rm pl}^2(r)}{\omega_{\rm ph}^2(r)}}\,.
\end{equation}
Thus, the real part of the QNMs and the shadow radius are related by a simple equation taking into account plasma effects
\begin{equation}
\omega_{\mathcal{\Re}} = \lim_{l \gg 1} \frac{l}{R_S}\sqrt{1-\frac{\omega_{\rm pl}^2(r)}{\omega_{\rm ph}^2(r)}},
\end{equation}
which is accurate only in the eikonal limit having large values of $l$. Nevertheless, deviations from this behavior may arise in modified theories of gravity, such as Einstein-Lovelock theory, where the link between null geodesics and QNMs can be disrupted \cite{Konoplya:2017wot,Wei:2019jve}. In addition, we can also express the Lyapunov exponent of the photon sphere in the following form \cite{Wei:2019jve}
\begin{align}
\mathsf{\lambda} = \sqrt{\frac{f(r_{\mathrm{ps}})\left[2f(r_{\mathrm{ps}}) - r_{\mathrm{ps}}^2 f''(r_{\mathrm{ps}})\right]}{2r_{\mathrm{ps}}^2}}\,.
\end{align}
In the case of a black hole with a topological defect, the Lyapunov exponent of the photon sphere for a homogeneous plasma medium ($\omega^2_{\rm pl}(r)=\text{const}$) is presented in Appendix \ref{Ap1}. When $k=0$, the Lyapunov exponent of the photon sphere reduces to the Schwarzschild case and is given as follows ($\omega^2_{\rm pl}(r)=\text{const}$)  \cite{Umarov:2025btg}
\begin{equation}
\lambda = \frac{2}{M} \sqrt{ \frac{ \left(1 - \eta\right)^2 \left( \sqrt{9 - 8\eta} - 1 \right) }{ \left( 3 - 4\eta + \sqrt{9 - 8\eta} \right)^3 } }, \quad \eta \equiv \frac{\omega_{\rm pl}^2}{\omega_0^2}\,,
\end{equation}
where, $\omega_0$ is the frequency measured by the observer at infinity.
\begin{figure}
    \centering
    \includegraphics[width=1\linewidth]{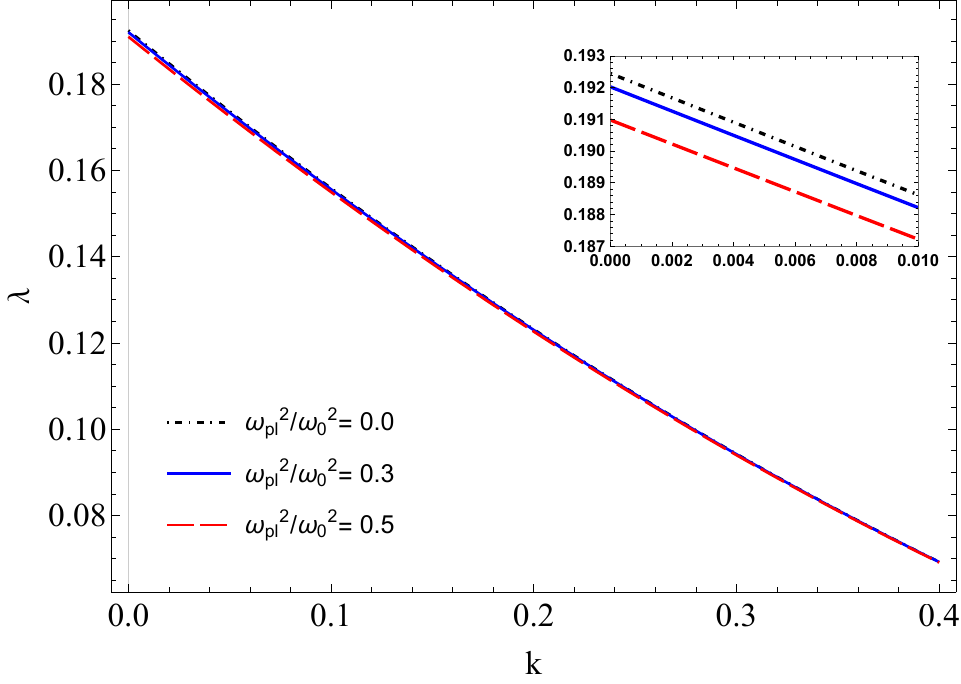}
    \caption{Dependence of the Lyapunov exponent of the photon sphere on the black hole parameter $k$ with a topological defect (M=1).}
    \label{fig:placeholder}\label{fig1}
\end{figure}

Fig. \ref{fig1} shows the dependence of the Lyapunov exponent of the photon sphere on the black hole parameter $k$ with a topological defect for different values of the homogeneous plasma frequency. It can be seen from the figure that the Lyapunov exponent depends very weakly on the plasma frequency however, as the plasma frequency increases, its value decreases. On the other hand, an increase in the parameter $k$ also leads to a decrease in the Lyapunov exponent.

\subsection{Inhomogeneous Plasma Profiles}

In order to investigate  inhomogeneous plasma profiles, we considered two models: the SIS and the non-NSIS. The SIS is a spherical cloud of gas with a single feature of density up to infinity at its center. The density distribution of a SIS is given by~\cite{Bisnovatyi:2010a, Rogers:2015dla}
\begin{align}
\rho(r)=\frac{\sigma_\nu^2}{2 \pi r^2}\, ,
\end{align}
where $\sigma_\nu^2$ refers to a one-dimensional velocity dispersion. The plasma concentration admits the following analytic expression~\cite{Bisnovatyi:2010a, Rogers:2015dla, Umarov:2025btg}
 \begin{equation}
     N(r)=\frac{\rho(r)}{am_{\rm p}}=\frac{\xi}{r^2}\, , \quad \xi=\frac{\sigma^2_\nu}{2\pi am_{\rm p}}\,,
     \label{24}
 \end{equation}
where $m_{\rm p}$ denotes the proton mass and $a$ is a dimensionless constant typically associated with the dark matter component of the Universe. Employing the expression for the plasma frequency, we obtain the following:
\begin{equation}
\omega_{\rm pl}^2=K_eN(r)=\frac{K_e\sigma_\nu^2}{2\pi a m_{\rm p}r^2}\, .
\end{equation}
where $K_e=4\pi e^2/m_e$, $m_e$ and $e$ are the electron mass and charge, respectively.

The NSIS is a more reasonable and physically motivated model for the analysis. Unlike the SIS, in this model the singularity is bounded by a finite core at the center of the gas cloud, and therefore the density distribution is defined as

\begin{equation}
\rho(r)=\frac{\sigma^2_\nu}{2\pi(r^2+r^2_c)}=\frac{\rho_0}{1+\frac{r^2}{r^2_c}}; \quad \rho_0=\frac{\sigma^2_\nu}{2\pi r^2_c}\,\label{2o}
\end{equation}
here, the core radius is represented by $r_c$. The plasma concentration for NSIS using Eq.~(\ref{24}) becomes
\begin{equation}
    N(r)=\frac{\xi}{r^2+r_c^2}\,.\label{2a}
\end{equation}
We compute the plasma frequency from expressions~(\ref{2o}) and (\ref{2a}) as follows
\begin{align}
\omega_{\rm pl}^2=\frac{K_e\sigma_\nu^2}{2\pi am_{\rm p}(r^2+r^2_c)}\, .
\end{align}

\section{QNMs of scalar field in plasma medium \label{sec:scalar_qnm}}
In studying the QNMs of a scalar field around a black hole, it is important to account for the influence of the surrounding environment. In this section, we model the plasma through an additional potential $\kappa N(r) \Phi^2,$ where \(N(r)\) is the radial plasma density and \(\kappa\) is the coupling constant. This formulation captures the effect of the plasma density on the scalar field, superimposed on the geometry of the black hole, which may include a topological defect. Using the square of the field \(\Phi^2\) ensures the symmetry \(\Phi \to -\Phi\) and preserves the linearity of the equation for small perturbations, which is necessary for a proper analysis of the QNMs. 

This approach is physically justified by the Klein–Gordon equation in curved spacetime and the principle of energy conservation: the plasma introduces a potential term that modifies the effective potential and the oscillation spectrum. The model allows one to clearly highlight the influence of plasma distribution on the frequencies and damping of QNMs, provides a convenient starting point for comparisons with more complex environmental models, and enables a systematic study of the effect of the surrounding medium on the dynamics of the scalar field.

Thus, the Klein–Gordon action for a real perturbed scalar field propagating in a plasma medium can be written as follows (here we shall introduce temporary the speed of light $c$ and Planck constant $\hbar$) \cite{Gelmini:2020xir}
 \begin{equation}
S = \int_M d^4x \, \sqrt{-g} \left( -\frac{1}{2} g^{\mu\nu} \nabla_\mu \Phi \nabla_\nu \Phi - \frac{m^2c^2}{2\hbar^2}  \Phi^2 +\kappa N(r)\Phi^2\right)\,.\label{10}
\end{equation}
On the other hand, from the Klein–Gordon action Eq. (\ref{10}), we can derive the Klein–Gordon equation in curved spacetime e taking into account the presence of plasma
\begin{align}
    \big(\partial_\mu\partial^\mu+m^2-2\chi(r)\big)\Phi=0\,,
\end{align}
where, $\chi(r)=\kappa N(r)$ characterizes the effects of the plasma medium.

Involving a separation of variables the function for the scalar field is given in terms of the spherical harmonics
\begin{equation}
\Phi(t, r, \theta, \phi) = \frac{1}{r} \, e^{-i \omega t} \, Y_{l}( \theta,\phi) \, \Psi(r).
\end{equation}
Here, \(l = 0,1,2,\ldots\) denotes the multipole number. After performing a separation of variables, it can be shown that the equation governing field perturbations in the black-hole spacetime, in the presence of a plasma medium, reduces to a Schrödinger-like wave equation.

\begin{equation}
    \frac{d^2\Psi}{dr^2_\star} + \big( \omega^2 - V_S(r) \big)\Psi = 0,
\end{equation}
 where we have used the relation
 \begin{equation}
     dr_\star=\frac{dr}{f(r)}\,,
 \end{equation}
 and the effective potential $V_S$ for the scalar field in the presence of plasma is obtained as follows
 \begin{equation}
     V_{S}(r)=f\bigg[\frac{l(l+1)}{r^2}+\frac{f'}{r}+m^2-2\chi(r)\bigg]\,.
 \end{equation}
where prime denotes the partial derivative with respect to radial coordinate as $f' \equiv\partial f/\partial r$. Under the positive real part, QNMs, by definition, satisfy the following boundary condition
 \begin{equation}
\Psi(r_\star) = C_{\pm} \exp\left(\pm i \omega r_\star\right),\quad r\rightarrow \pm \infty\,.
 \end{equation}
 where \(\omega\) can be written in terms of the real and imaginary part i.e., \(\omega = \omega_{\Re} - i\omega_{\Im}\). In other words, we have the real oscillation frequency and the imaginary part which is proportional to the decay rate of a given mode. The corresponding perturbation of the scalar field in the case of a black hole with a topological defect has the following effective potential
 \begin{equation}
    V_S(r)= \left(1-k-\frac{2 M}{r}\right) \left(\frac{l (l+1)}{r^2}+\frac{2 M}{r^3}+m ^2-2 \chi(r)\right)\,.
 \end{equation}
 \begin{figure*}
   \centering
{\includegraphics[width=0.326\textwidth]{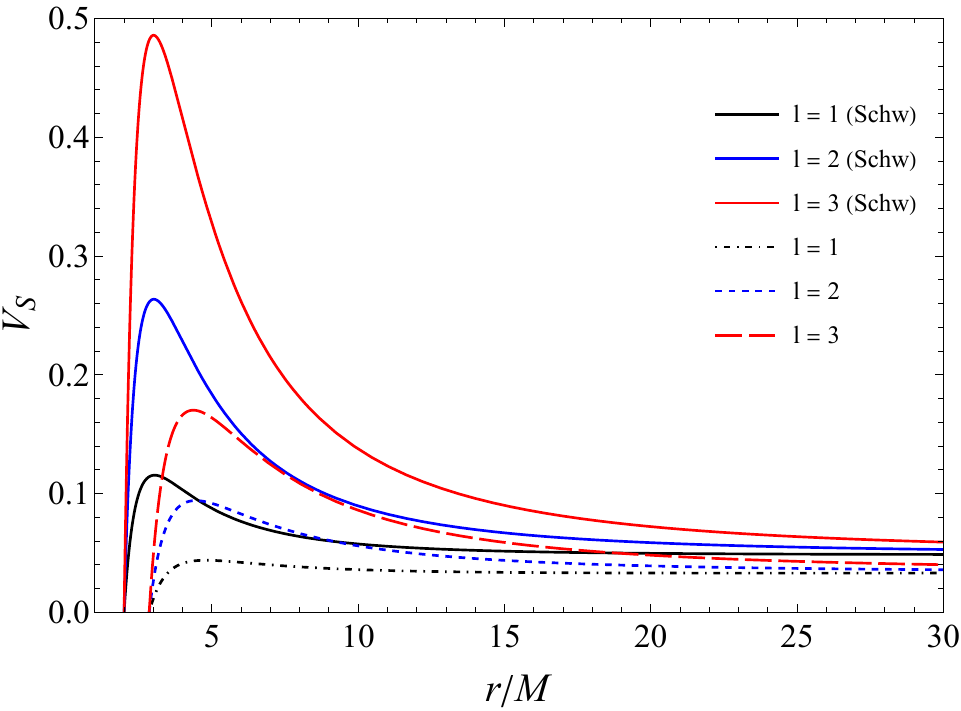}}\hfill
{\includegraphics[width=0.326\textwidth]{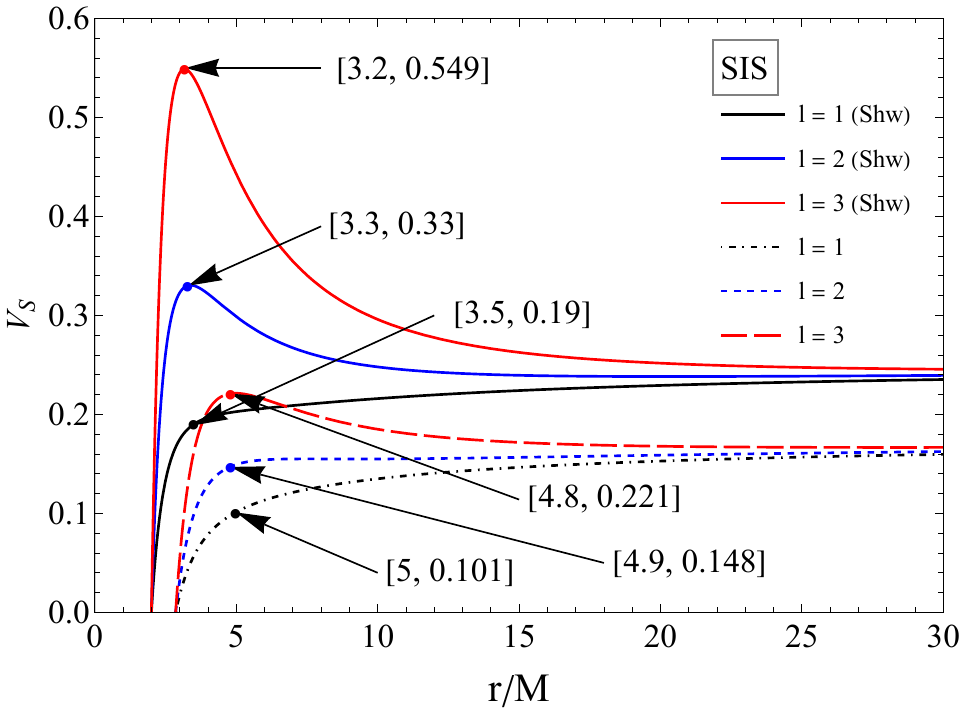}}\hfill
 {\includegraphics[width=0.326\textwidth]{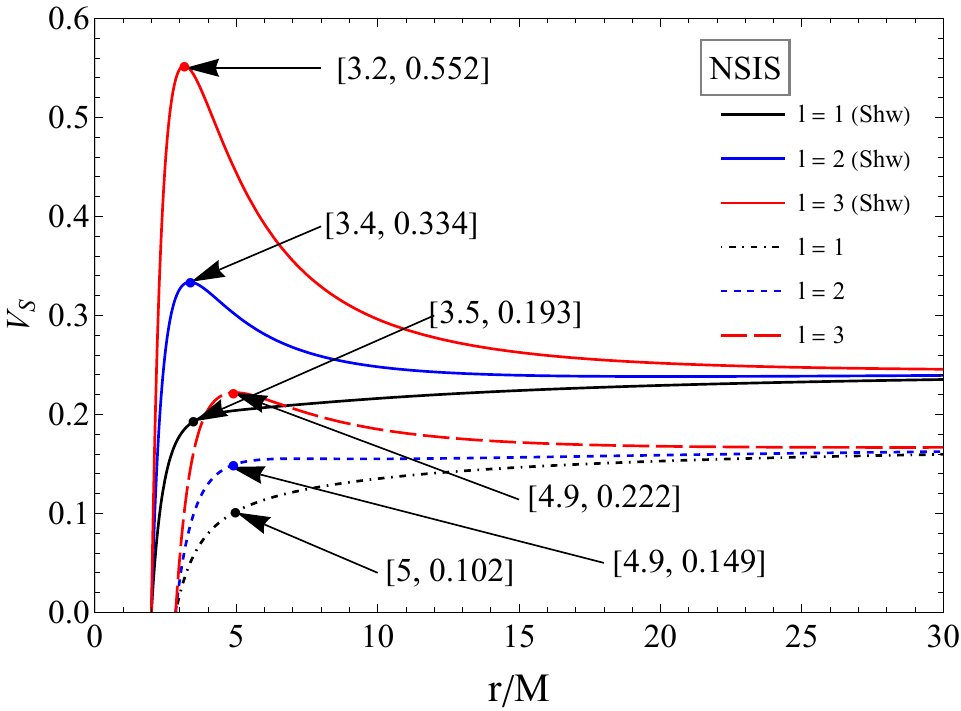}}\hfill
\caption{Effective potentials of scalar perturbations for different plasma profiles. The left panel corresponds to a homogeneous plasma medium, while the middle and right panels correspond to inhomogeneous plasma models, namely the SIS and NSIS models, respectively. The parameters are fixed as $m=0.5$, $\chi=0.1$, $r_c=3$ and  $\kappa \xi=0.1.$
}
    \label{s8}
\end{figure*}
 
 \begin{figure*}[ht]
   \centering
   {\includegraphics[width=0.4\textwidth]{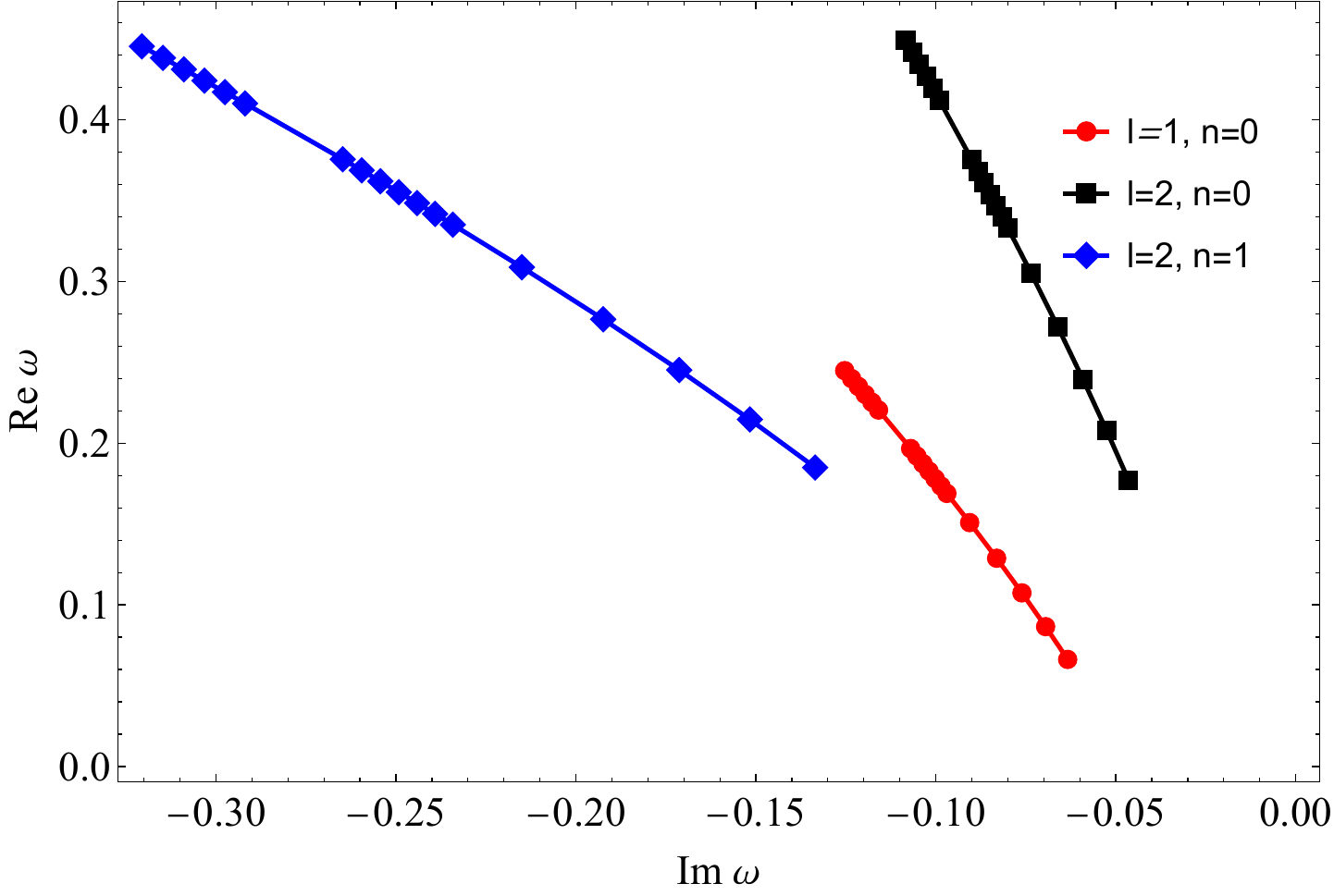}}\hspace{10mm}
   {\includegraphics[width=0.4\textwidth]{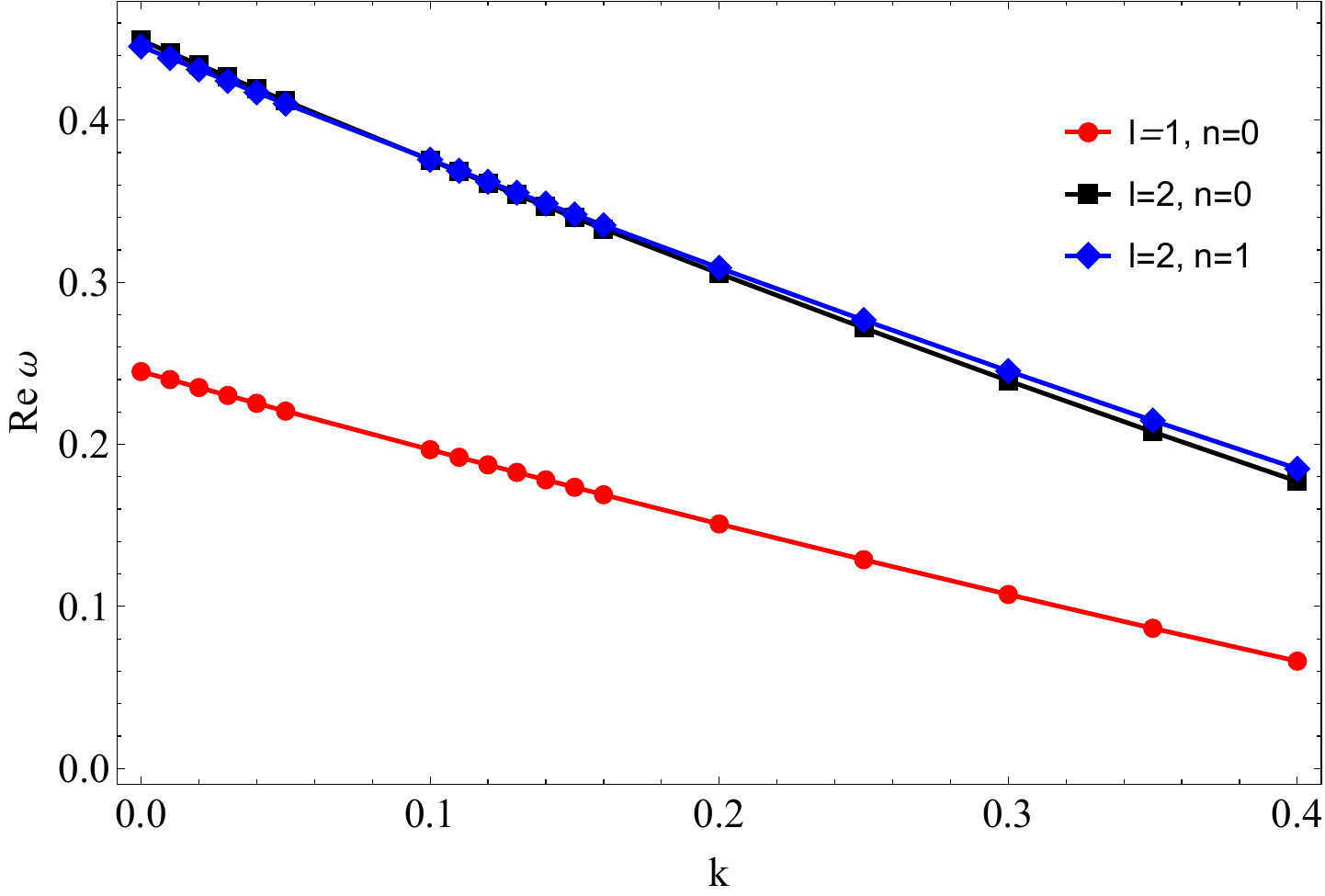}}
   \caption{Left panel: Dependence of the real part of the QNM frequencies on the imaginary part for the scalar field.
Right panel: Dependence of the real part of the QNM frequencies on the parameter $k$ for a black hole with a topological defect in a homogeneous plasma medium. The parameters were chosen as $m=0.3$, $\chi=0.1$, and $k\in[0,0.4]$.}
   \label{fig3}
\end{figure*}
\begin{figure*}[ht]
   \centering
   {\includegraphics[width=0.4\textwidth]{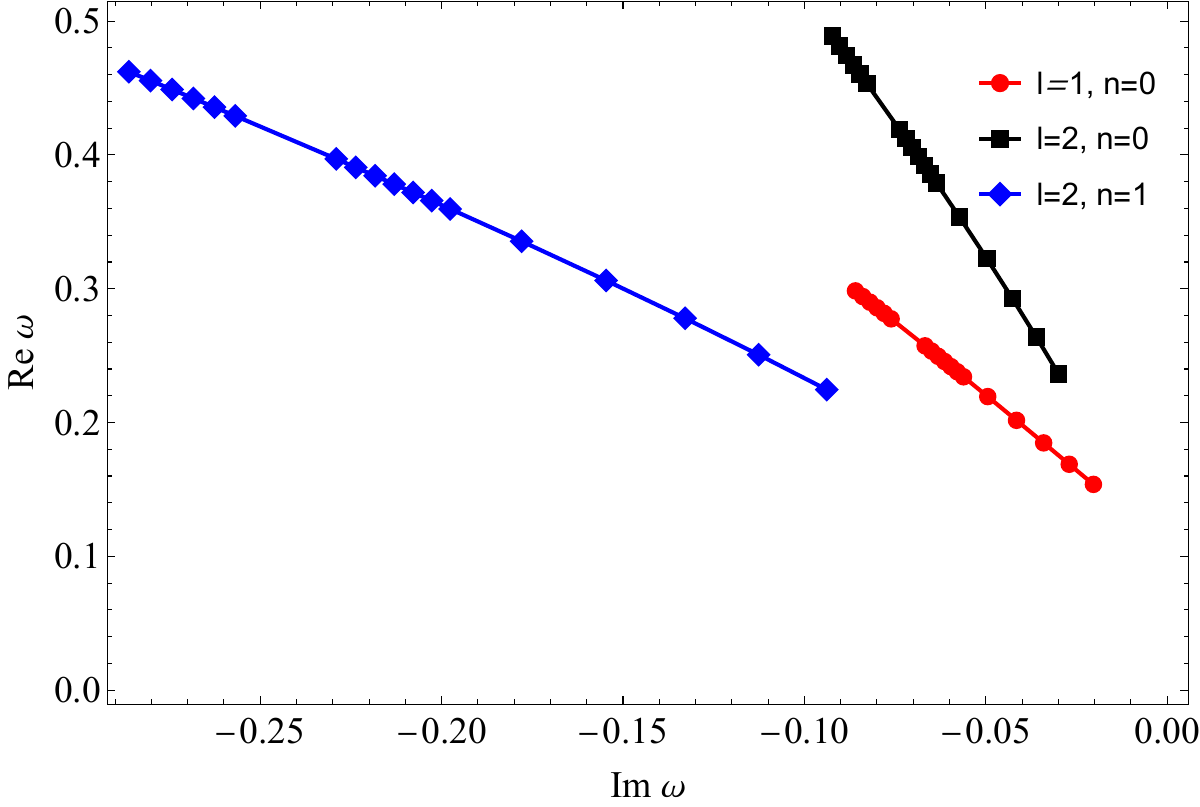}}\hspace{10mm}
   {\includegraphics[width=0.4\textwidth]{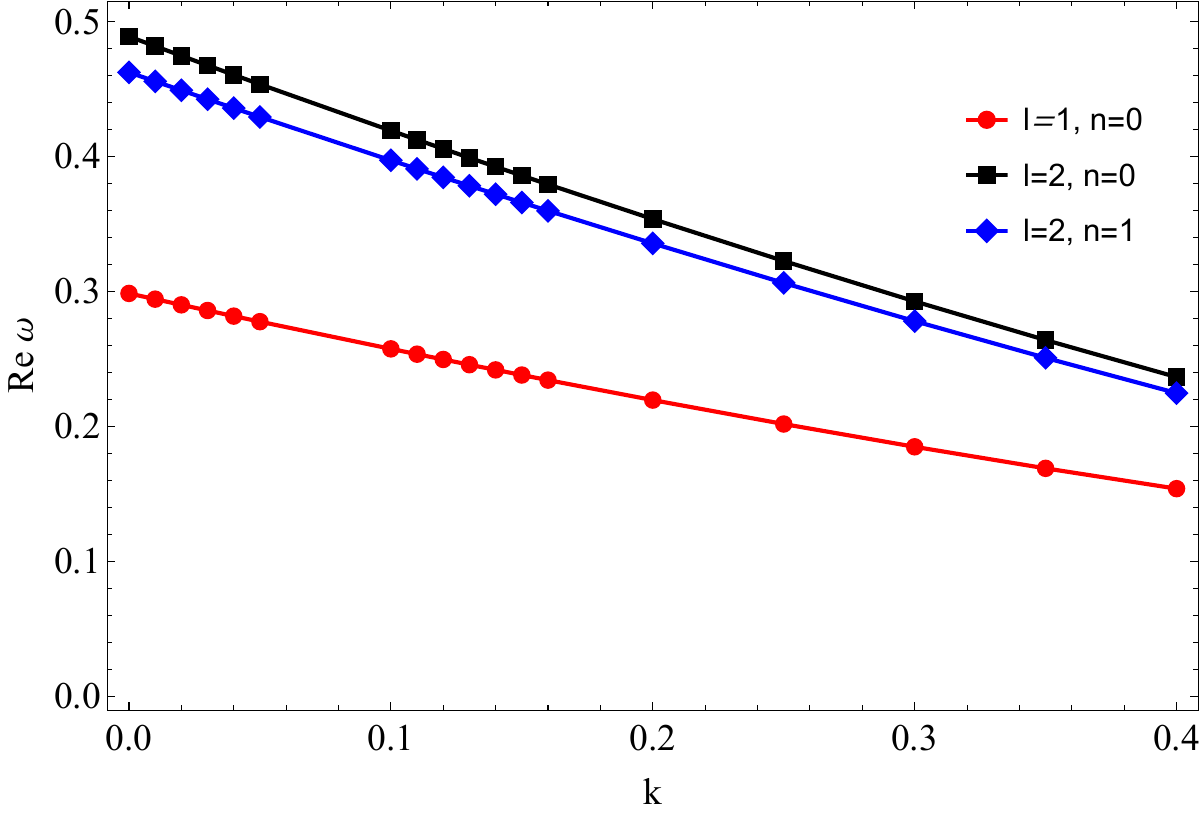}}
   \caption{Left panel: Dependence of the real part of the QNM frequencies on their imaginary part for a scalar field. Right panel: Dependence of the real part of the QNM frequencies on the parameter $k$ for a black hole with a topological defect in an inhomogeneous plasma described by the SIS model. The parameters are fixed as $m = 0.2$, $\kappa\xi = 0.1$, and $k \in [0, 0.4]$.}
   \label{fig4}
\end{figure*}
\begin{figure*}[ht]
   \centering
   {\includegraphics[width=0.4\textwidth]{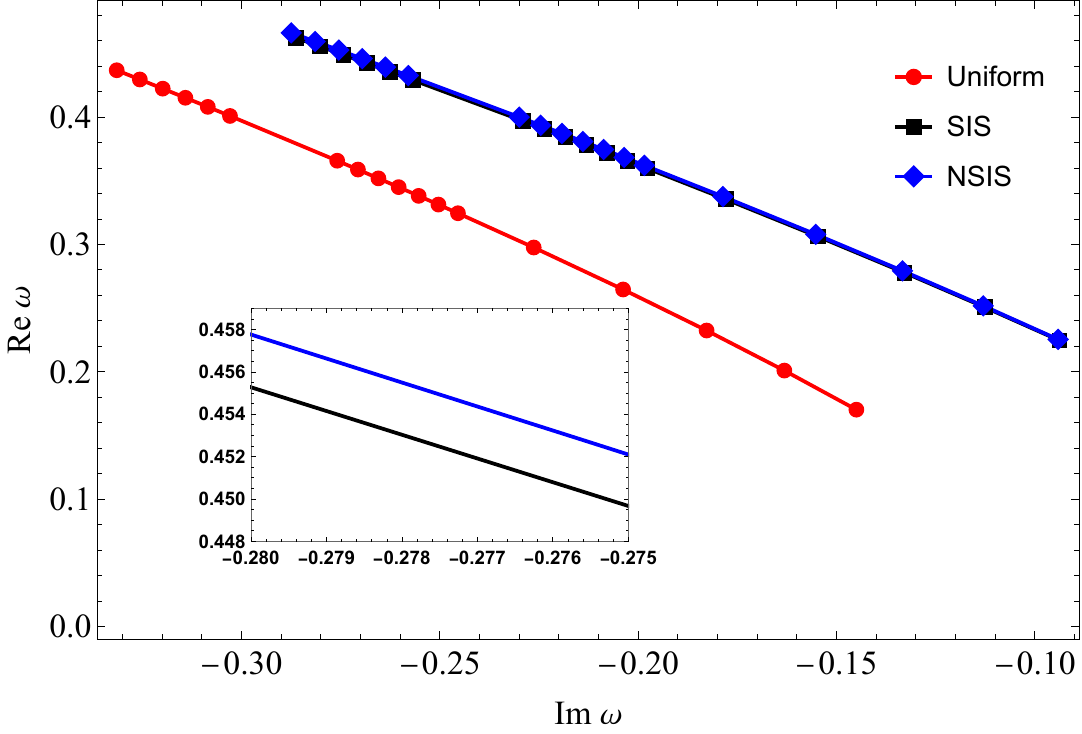}}\hspace{10mm}
   {\includegraphics[width=0.4\textwidth]{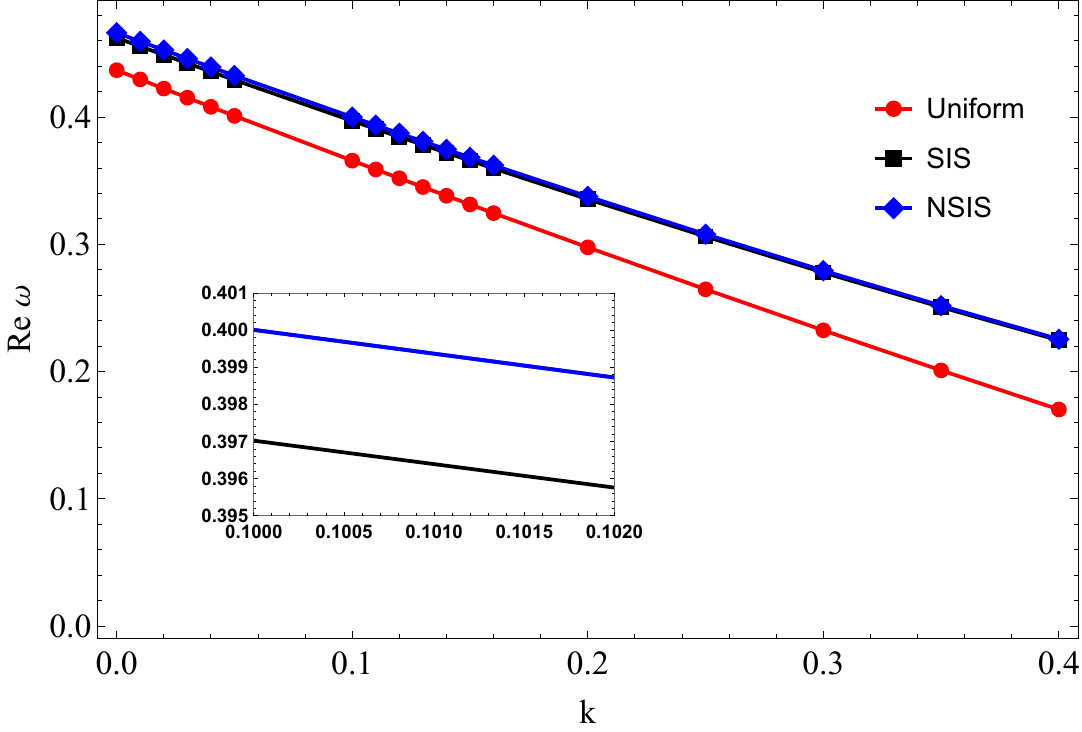}}
   \caption{Left panel: Dependence of the real part of scalar QNM frequencies on their imaginary part for different plasma models (Uniform, SIS, NSIS). 
Right panel: Real part of the QNM frequency as a function of $k$ for a black hole with a topological defect, shown for the same plasma models (Uniform, SIS, NSIS).  The parameters are fixed as  $m=0.2$, $\kappa \xi=0.1$, $r_c=3$, $l=2, n=1$ and $k\in[0,0.4]$.}
   \label{fig5}
\end{figure*}

In Fig.~\ref{s8}, the effective potentials of scalar perturbations $V_S$ are shown as functions of the dimensionless radial coordinate $r/M$ for different values of the multipole number $l$ in a plasma medium for the vacuum case, see \cite{Gogoi:2022wyv,Konoplya:2003ii,Konoplya:2017wot} The solid curves correspond to the Schwarzschild case ($k = 0$), whereas the dashed curves represent a black hole with a topological effect ($k \neq 0$).

In the left panel, the case of homogeneous plasma ($\chi = \mathrm{const}$) is shown. It is clearly seen that for fixed values of the black hole parameter $k$, the scalar field mass $m$, and the plasma parameter $\chi$, an increase in the multipole number $l$ leads to an increase in the height of the potential barrier. This behavior is expected and is associated with the contribution of the centrifugal term $\sim l(l+1)/r^2$, which enhances the effective barrier for larger $l$. 

A comparison of the solid and dashed curves shows that the topological parameter $k$ effectively weakens the gravitational contribution to the formation of the potential barrier. As a result, the maximum of the effective potential decreases, which may lead to a reduction in the real part of the QNMs frequencies, as well as to a modification of the damping rate of the perturbations. Physically, this means that the presence of a topological effect modifies the spacetime geometry in such a way that the gravitational and centrifugal contributions to the effective potential become less pronounced compared to the pure Schwarzschild case.

In the middle and right panels, the case of inhomogeneous plasma is shown for two models SIS and NSIS. The transition from homogeneous plasma to these models led to quantitative changes in the height and position of the effective potential’s maximum. In the SIS and NSIS models, the potential maximum was slightly shifted, and its height increased compared to the homogeneous case. This was explained by the radial dependence of the plasma frequency, which modified the effective dispersion of the scalar field. The differences between SIS and NSIS in barrier heights and maximum positions were minor and were caused by differences in the plasma density profiles. The qualitative shape of the potential remained the same a barrier with a maximum near the photon sphere.

Having obtained the expression for the effective potential, we employed the WKB approach to compute the QNMs frequencies. The WKB method was based on an analogy with the problem of wave scattering near the peak of a potential barrier in quantum mechanics, where the frequency $\omega$ played the role of energy. This approach was originally proposed by Schutz and Will \cite{1985ApJ...291L..33S} and later extended to third order by Iyer and Will \cite{Iyer:1986np}. In the present work, we used the sixth-order WKB approximation developed by Konoplya \cite{Konoplya:2003ii} to calculate the QNMs. 

In Tables \ref{tab-1}–\ref{tab-3}, the real and imaginary parts of the QNM frequencies of a scalar field in the background of a black hole with a topological defect and in the Schwarzschild background were presented for various plasma models. The calculations were performed within the framework of the third- and sixth-order WKB approximations.

Fig. \ref{fig3}  shows the spectral characteristics of the QNMs of a scalar field in the background of a black hole with a topological defect in a homogeneous plasma medium. The left panel demonstrated the dependence of the real part of the frequency on its imaginary part for different multipole numbers and overtones. For each pair $(l, n),$ an almost linear correlation between $\text{Re} \omega$ and $\text {Im} \omega$ was observed. Modes with larger $l$ were located higher, meaning they possessed a larger real part of the frequency. This was consistent with the fact that increasing $l$ raised the effective potential barrier, leading to higher oscillation frequencies. The right panel showed the dependence of the real part of the frequency on the topological parameter $k$. In the considered range $k \in [0, 0.4]$, a monotonic decrease of $\mathrm{Re}\,\omega$ was observed as $k$ increased. This indicated that strengthening the topological defect led to a decrease in the oscillation frequency compared to the Schwarzschild background \cite{Alves:2025erh}.

On the other hand, for all values of $k$, the following inequality held:
\[
\mathrm{Re}\,\omega(l=2,n=0)
>
\mathrm{Re}\,\omega(l=2,n=1)
>
\mathrm{Re}\,\omega(l=1,n=0).
\]
This reflected a stable structure of the spectrum determined by the shape of the effective potential.

Table \ref{tab-1} presented the real and imaginary parts of the QNM frequencies of a scalar field in the background of a black hole with a topological defect, as well as in the Schwarzschild background, both immersed in a homogeneous plasma, for various sets of parameters $(m,k,\chi)$. A comparison of the tables reveals two main differences. First, the $(0,0)$ mode is present only in the case of a black hole with a topological defect, while it is absent in the Schwarzschild background. Second, for identical modes, the corresponding frequencies in the defect case are systematically lower. Therefore, the presence of the defect leads to a simultaneous reduction of both the oscillation frequency and the damping rate.

In contrast, a comparison with the vacuum results reported in \cite{Konoplya:2003ii} shows a qualitatively different behavior in the presence of a homogeneous plasma. In particular, the real part of the frequency $\mathrm{Re},\omega$ decreases, whereas the magnitude of the imaginary part $|\mathrm{Im},\omega|$ increases. This indicates that the plasma not only slows down wave propagation but also enhances the dissipation of perturbations by increasing their tunneling through the effective potential barrier.

Fig. \ref{fig4} presents the spectral characteristics of the scalar QNMs of a black hole with a topological defect in the presence of an inhomogeneous plasma, whose density distribution is described by the SIS model. The left panel shows the dependence of the real part of the frequency $\mathrm{Re}\,\omega$ on the imaginary part $\mathrm{Im}\,\omega$ for different multipole numbers $l$ and overtones $n$. As in the case of homogeneous plasma (Fig. \ref{fig3}), an almost linear correlation between these quantities is observed. Modes with larger values of the multipole number $l$ are characterized by a larger real part of the frequency, which is due to the increase in the height of the effective potential barrier. The right panel shows the dependence of the real part of the frequency on the topological parameter $k$ in the range $k \in [0, 0.4]$. The qualitative behavior of the frequency is fully analogous to the case of homogeneous plasma presented in Fig. \ref{fig3} increasing the parameter $k$, which characterizes the strength of the topological defect, leads to a monotonic decrease of $\mathrm{Re}\,\omega$.

Fig. \ref{fig5}  presents a comparative analysis of the influence of the type of plasma environment on the QNM spectrum. Three models are considered: homogeneous plasma, SIS, and NSIS.

The left panel showed the dependence of the real part of the frequency $\mathrm{Re}\omega  $ on its imaginary part $\mathrm{Im}\omega$ for three plasma models. It was easy to notice that in the NSIS model the oscillation frequency $\mathrm{Re}\omega $ took slightly larger values than in the case of homogeneous plasma. At the same time, for the NSIS and SIS models the difference between the values of $\mathrm{Re}\omega  $ and $\mathrm{Im}\omega  $ remained very small, indicating that the spectral characteristics of these two plasma profiles were quite similar.

The right panel showed the dependence of the real part of the frequency $\mathrm{Re}\omega$ on the topological parameter  $k$ for the same plasma models. It was easy to notice that an increase in the parameter  $k$  led to a monotonic decrease in $\mathrm{Re}\omega  $. However, the magnitude of this effect depended on the plasma profile: the largest values of the frequency were observed in the NSIS model, followed by the SIS model, while in the case of homogeneous plasma the values of $\mathrm{Re}\omega $ were the smallest.

\begin{widetext}

\begin{table}[h]
\centering
\caption{The real and imaginary parts of the QNMs of a scalar field in the background of a black hole with a topological defect in a uniform plasma medium. The upper tables correspond to the black hole with a topological defect, while the lower tables correspond to the Schwarzschild case.}
\label{tab:QNM_scalar_k0_chi0}
\renewcommand{\arraystretch}{1.3}
\begin{tabular}{|c|c|c|}
\hline
\multicolumn{3}{|c|}{$m=0.3$, $k=0.1$, $\chi=0.1$} \\
\hline
$(l,n)$ & 3rd order WKB & 6th order WKB \\
\hline
$(0,0)$ & $0.0368 - 0.1668 i$ & $0.0298 - 0.2032 i$ \\
$(1,0)$ & $0.1957 - 0.1074 i$ & $0.1967 - 0.1070 i$ \\
$(1,1)$ & $0.2188 - 0.3044 i$ & $0.1913 - 0.2951 i$ \\
$(2,0)$ & $0.3751 - 0.0900 i$ & $0.3755 - 0.0900 i$ \\
$(2,1)$ & $0.3751 - 0.2658 i$ & $0.3757 - 0.2648 i$ \\
$(2,2)$ & $0.3680 - 0.4364 i$ & $0.3608 - 0.4362 i$ \\
\hline
\end{tabular}
\begin{tabular}{|c|c|c|}
\hline
\multicolumn{3}{|c|}{$m=0.4$, $k=0.2$, $\chi=0.15$} \\
\hline
$(l,n)$ & 3rd order WKB & 6th order WKB \\
\hline
$(0,0)$ & $0.0487 - 0.1685 i$ & $0.0485 - 0.1686 i$ \\
$(1,0)$ & $0.1582 - 0.1406 i$ & $0.1593 - 0.1400 i$ \\
$(1,1)$ & $0.2046 - 0.3156 i$ & $0.2026 - 0.3120 i$ \\
$(2,0)$ & $0.2985 - 0.1110 i$ & $0.2989 - 0.1109 i$ \\
$(2,1)$ & $0.3235 - 0.3006 i$ & $0.3240 - 0.3004 i$ \\
$(2,2)$ & $0.3303 - 0.4687 i$ & $0.3294 - 0.4725 i$ \\
\hline
\end{tabular}
\begin{tabular}{|c|c|c|}
\hline
\multicolumn{3}{|c|}{$m=0.7$, $k=0.3$, $\chi=0.3
$} \\
\hline
$(l,n)$ & 3rd order WKB & 6th order WKB \\
\hline
$(0,0)$ & $0.0753 - 0.1563i$ & $0.0776 -0.1529 i$ \\
$(1,0)$ & $0.1427 - 0.1467 i$ & $0.1439 - 0.1460 i$ \\
$(1,1)$ & $0.1871 - 0.3168 i$ & $0.1887 - 0.3164 i$ \\
$(2,0)$ & $0.2459 - 0.1186 i$ & $0.2463 - 0.1185 i$ \\
$(2,1)$ & $0.2731 - 0.3140i$ & $0.2737 - 0.3151 i$ \\
$(2,2)$ & $0.2785 - 0.4923 i$ & $0.2863 - 0.5034 i$ \\
\hline
\end{tabular}\label{tab-1}

\begin{tabular}{|c|c|c|}
\hline
\multicolumn{3}{|c|}{Shw, $m=0.3$, $\chi=0.1$ }\\
\hline
$(l,n)$ & 3rd order WKB & 6th order WKB \\
\hline
$(0,0)$ & $-$ & $-$ \\
$(1,0)$ & $0.2434 -0.1258 i$ & $0.2449 -0.1253 i$ \\
$(1,1)$ & $0.2598-0.3583 i$ & $0.2378-0.3498 i$ \\
$(2,0)$ & $0.4489 -0.1084 i$ & $0.4493 -0.1084 i$ \\
$(2,1)$ & $0.4448 -0.3216 i$ & $0.4455 -0.3206 i$ \\
$(2,2)$ & $0.3680 - 0.4364 i$ & $0.4235 -0.5325 i$ \\
\hline
\end{tabular}
\begin{tabular}{|c|c|c|}
\hline
\multicolumn{3}{|c|}{Shw, $m=0.4$, $\chi=0.15$ }\\
\hline
$(l,n)$ & 3rd order WKB & 6th order WKB \\
\hline
$(0,0)$ & $-$ & $-$ \\
$(1,0)$ & $0.2308 -0.1328 i$ & $0.2322 -0.1323 i$ \\
$(1,1)$ & $0.2598 -0.3761 i$ & $0.2184 -0.3668 i$ \\
$(2,0)$ & $0.4397 -0.1115 i$ & $0.4401 -0.1114 i$ \\
$(2,1)$ & $0.4397 -0.3286 i$ & $0.4403 -0.3272 i$ \\
$(2,2)$ & $0.4303 -0.5395 i$ & $0.4187 -0.5398 i$ \\
\hline
\end{tabular}
\begin{tabular}{|c|c|c|}
\hline
\multicolumn{3}{|c|}{Shw, $m=0.7$, $\chi=0.3
$} \\
\hline
$(l,n)$ & 3rd order WKB & 6th order WKB \\
\hline
$(0,0)$ & $-$ & $-$ \\
$(1,0)$ & $0.2434 -0.1258 i$ & $0.2449 -0.1253 i$ \\
$(1,1)$ & $0.2598 -0.3583i$ & $0.2378 -0.3498 i$ \\
$(2,0)$ & $0.4489 -0.1084 i$ & $0.4493 -0.1084 i$ \\
$(2,1)$ & $0.4448 -0.3216i$ & $0.4455 -0.3206 i$ \\
$(2,2)$ & $0.4306 -0.5305 i$ & $0.4235 -0.5325 i$ \\
\hline
\end{tabular}\label{tab-1}
\end{table}
\begin{table}[h]
\centering
\caption{The real and imaginary parts of the QNMs of a scalar field in the background of a black hole with a topological defect in an inhomogeneous plasma medium described by the SIS model. The upper tables correspond to the black hole with a topological defect, while the lower tables correspond to the Schwarzschild case.
}
\renewcommand{\arraystretch}{1.3}
\label{tab:QNM_scalar_k0_chi0}
\begin{tabular}{|c|c|c|}
\hline
\multicolumn{3}{|c|}{$m=0.1$, $k=0.1$, $\chi=0.1$} \\
\hline
$(l,n)$ & 3rd order WKB & 6th order WKB \\
\hline
$(0,0)$ & $0.0595 - 0.1002 i$ & $0.0514 - 0.0889 i$ \\
$(1,0)$ & $0.2408 - 0.0763 i$ & $0.2422 - 0.0762 i$ \\
$(1,1)$ & $0.2125 - 0.2456 i$ & $0.2145 - 0.2445 i$ \\
$(2,0)$ & $0.4084 - 0.0772 i$ & $0.4087 - 0.0772 i$ \\
$(2,1)$ & $0.3911 - 0.2367 i$ & $0.3916 - 0.2366 i$ \\
$(2,2)$ & $0.3643 - 0.4046 i$ & $0.3633 - 0.4084 i$ \\
\hline
\end{tabular}
\begin{tabular}{|c|c|c|}
\hline
\multicolumn{3}{|c|}{$m=0.2$, $k=0.2$, $\chi=0.4$} \\
\hline
$(l,n)$ & 3rd order WKB & 6th order WKB \\
\hline
$(0,0)$ & $0.1817 - 0.0048 i$ & $0.1817 - 0.0047 i$ \\
$(1,0)$ & $0.1880 - 0.0451 i$ & $0.1898 - 0.0450 i$ \\
$(1,1)$ & $0.1444 - 0.2016 i$ & $0.1509 - 0.1762 i$ \\
$(2,0)$ & $0.3367 - 0.0568 i$ & $0.3370 - 0.0568 i$ \\
$(2,1)$ & $0.3169 - 0.1778 i$ & $0.3175 - 0.1776 i$ \\
$(2,2)$ & $0.2896 - 0.3125 i$ & $0.2897 - 0.3139 i$ \\
\hline
\end{tabular}
\begin{tabular}{|c|c|c|}
\hline
\multicolumn{3}{|c|}{$m=0.1$, $k=0.4$, $\chi=0.2
$} \\
\hline
$(l,n)$ & 3rd order WKB & 6th order WKB \\
\hline
$(0,0)$ & $0.0781 - 0.0020i$ & $0.0781 -0.0020 i$ \\
$(1,0)$ & $0.1262 - 0.0314 i$ & $0.1266 - 0.0314 i$ \\
$(1,1)$ & $0.1103 - 0.1044 i$ & $0.1114 - 0.1035 i$ \\
$(2,0)$ & $0.2195 - 0.0336 i$ & $0.2196 - 0.0336 i$ \\
$(2,1)$ & $0.2120 - 0.1027i$ & $0.2122 - 0.1026 i$ \\
$(2,2)$ & $0.2001 - 0.1759 i$ & $0.1996 - 0.1768 i$ \\
\hline
\end{tabular}\label{tab-2}

\begin{tabular}{|c|c|c|}
\hline
\multicolumn{3}{|c|}{Shw, $m=0.1$, $\chi=0.1$} \\
\hline
$(l,n)$ & 3rd order WKB & 6th order WKB \\
\hline
$(0,0)$ & $0.0759 -0.1222 i$ & $0.0707 -0.1073 i$ \\
$(1,0)$ & $0.2825 -0.0953 i$ & $0.2846 -0.0950 i$ \\
$(1,1)$ & $0.2483 -0.3060 i$ & $0.2509 -0.3045 i$ \\
$(2,0)$ & $0.4787 -0.0957 i$ & $0.4791 -0.0957 i$ \\
$(2,1)$ & $0.4568 -0.2937 i$ & $0.4575 -0.2935 i$ \\
$(2,2)$ & $0.4234 -0.5020 i$ & $0.4223 -0.5074 i$ \\
\hline
\end{tabular}
\begin{tabular}{|c|c|c|}
\hline
\multicolumn{3}{|c|}{Shw, $m=0.2$, $\chi=0.4$} \\
\hline
$(l,n)$ & 3rd order WKB & 6th order WKB \\
\hline
$(0,0)$ & $-$ & $-$ \\
$(1,0)$ & $0.2532 -0.0831 i$ & $0.2566 -0.0826 i$ \\
$(1,1)$ & $0.2015 -0.3138 i$ & $0.2054 -0.2969 i$ \\
$(2,0)$ & $0.4649 -0.0919i$ & $0.4654 -0.0919 i$ \\
$(2,1)$ & $0.4361 -0.2867 i$ & $0.4371 -0.2864 i$ \\
$(2,2)$ & $0.3969 -0.4989 i$ & $0.3966 -0.5037 i$ \\
\hline
\end{tabular}
\begin{tabular}{|c|c|c|}
\hline
\multicolumn{3}{|c|}{Shw, $m=0.1$, $\chi=0.2
$} \\
\hline
$(l,n)$ & 3rd order WKB & 6th order WKB \\
\hline
$(0,0)$ & $0.0516 -0.1443i$ & $0.0271 -0.1500 i$ \\
$(1,0)$ & $0.2688 -0.0952 i$ & $0.2711 -0.0949 i$ \\
$(1,1)$ & $0.2336 -0.3080i$ & $0.2365 -0.3062 i$ \\
$(2,0)$ & $0.4708 -0.0957 i$ & $0.4713 -0.0957 i$ \\
$(2,1)$ & $0.4486 -0.2939i$ & $0.4493 -0.2936 i$ \\
$(2,2)$ & $0.4149 -0.5026 i$ & $0.4138 -0.5082 i$ \\
\hline
\end{tabular}\label{tab-2}
\end{table}
\begin{table}[h]
\centering
\caption{The real and imaginary parts of the QNMs of a scalar field in the background of a black hole with a topological defect in an inhomogeneous plasma medium described by the NSIS model. The upper tables correspond to the black hole with a topological defect, while the lower tables correspond to the Schwarzschild case.}
\renewcommand{\arraystretch}{1.3}
\label{tab:QNM_scalar_k0_chi0}
\begin{tabular}{|c|c|c|}
\hline
\multicolumn{3}{|c|}{$m=0.1$, $k=0.1$, $\chi=0.1$} \\
\hline
$(l,n)$ & 3rd order WKB & 6th order WKB \\
\hline
$(0,0)$ & $0.0770 - 0.0969 i$ & $0.0712 - 0.0932 i$ \\
$(1,0)$ & $0.2457 - 0.0772 i$ & $0.2471 - 0.0769 i$ \\
$(1,1)$ & $0.2185 - 0.2477 i$ & $0.2200 - 0.2463 i$ \\
$(2,0)$ & $0.4113 - 0.0775 i$ & $0.4116 - 0.0775 i$ \\
$(2,1)$ & $0.3941 - 0.2376 i$ & $0.3946 - 0.2375 i$ \\
$(2,2)$ & $0.3677 - 0.4061 i$ & $0.3665 - 0.4098 i$ \\
\hline
\end{tabular}
\begin{tabular}{|c|c|c|}
\hline
\multicolumn{3}{|c|}{$m=0.1$, $k=0.2$, $\chi=0.3$} \\
\hline
$(l,n)$ & 3rd order WKB & 6th order WKB \\
\hline
$(0,0)$ & $0.0488 - 0.0982 i$ & $0.0080 - 0.1552 i$ \\
$(1,0)$ & $0.1932 - 0.0614 i$ & $0.1945 - 0.0609 i$ \\
$(1,1)$ & $0.1712 - 0.1997 i$ & $0.1718 - 0.1977 i$ \\
$(2,0)$ & $0.3375 - 0.0614 i$ & $0.3377 - 0.0613 i$ \\
$(2,1)$ & $0.3239 - 0.1881 i$ & $0.3242 - 0.1879 i$ \\
$(2,2)$ & $0.3030 - 0.3217 i$ & $0.3020 - 0.3244 i$ \\
\hline
\end{tabular}
\begin{tabular}{|c|c|c|}
\hline
\multicolumn{3}{|c|}{$m=0$, $k=0.4$, $\chi=0.2
$} \\
\hline
$(l,n)$ & 3rd order WKB & 6th order WKB \\
\hline
$(0,0)$ & $0.0200 - 0.0567i$ & $0.0162 -0.0596 i$ \\
$(1,0)$ & $0.1227 - 0.0356 i$ & $0.1232 - 0.0355 i$ \\
$(1,1)$ & $0.1132 - 0.1106 i$ & $0.1138 - 0.1102 i$ \\
$(2,0)$ & $0.2171 - 0.0349 i$ & $0.2172 - 0.0349 i$ \\
$(2,1)$ & $0.2111 - 0.1061i$ & $0.2113 - 0.1060 i$ \\
$(2,2)$ & $0.2011 - 0.1797 i$ & $0.2007 - 0.1807 i$ \\
\hline
\end{tabular}\label{tab-3}

\begin{tabular}{|c|c|c|}
\hline
\multicolumn{3}{|c|}{Shw, $m=0.1$, $\chi=0.1$} \\
\hline
$(l,n)$ & 3rd order WKB & 6th order WKB \\
\hline
$(0,0)$ & $0.0974 -0.1191 i$ & $0.0925 -0.1146 i$ \\
$(1,0)$ & $0.2890 -0.0964 i$ & $0.2910 -0.0959 i$ \\
$(1,1)$ & $0.2565 -0.3086 i$ & $0.2581 -0.3072 i$ \\
$(2,0)$ & $0.4825 -0.0961 i$ & $0.4829 -0.0960 i$ \\
$(2,1)$ & $0.4608 -0.2948 i$ & $0.4615 -0.2946 i$ \\
$(2,2)$ & $0.4280 -0.5039 i$ & $0.4266 -0.5093 i$ \\
\hline
\end{tabular}
\begin{tabular}{|c|c|c|}
\hline
\multicolumn{3}{|c|}{Shw, $m=0.1$, $\chi=0.3$} \\
\hline
$(l,n)$ & 3rd order WKB & 6th order WKB \\
\hline
$(0,0)$ & $0.0846 -0.1394 i$ & $0.0773 -0.1430 i$ \\
$(1,0)$ & $0.2757 -0.0990i$ & $0.2777 -0.0980 i$ \\
$(1,1)$ & $0.2460 -0.3179 i$ & $0.2457 -0.3173 i$ \\
$(2,0)$ & $0.4747 -0.0969 i$ & $0.4751 -0.0968 i$ \\
$(2,1)$ & $0.4529 -0.2976 i$ & $0.4534 -0.2973 i$ \\
$(2,2)$ & $0.4207 -0.5090 i$ & $0.4190 -0.5151 i$ \\
\hline
\end{tabular}
\begin{tabular}{|c|c|c|}
\hline
\multicolumn{3}{|c|}{Shw, $m=0$, $\chi=0.2
$} \\
\hline
$(l,n)$ & 3rd order WKB & 6th order WKB \\
\hline
$(0,0)$ & $0.0882 -0.1344i$ & $0.0818 -0.1351 i$ \\
$(1,0)$ & $0.2780 -0.1005 i$ & $0.2800 -0.0997 i$ \\
$(1,1)$ & $0.2511 -0.3158 i$ & $0.2517 -0.3151i$ \\
$(2,0)$ & $0.4754 -0.0976 i$ & $0.4759 -0.0975 i$ \\
$(2,1)$ & $0.4554 -0.2985i$ & $0.4559 -0.2983 i$ \\
$(2,2)$ & $0.4243 -0.5083 i$ & $0.4226 -0.5142 i$ \\
\hline
\end{tabular}\label{tab-3}
\end{table}

\end{widetext}

\section{ PLASMA-PHOTON INTERACTION IN
CURVED SPACETIME: BASIC EQUATION\label{sec:em_plasma}}

A comprehensive analysis of electromagnetic wave propagation in cold plasma within curved spacetime was first presented in \cite{BreuerEhlers1981}, where a set of nonlinear equations describing the coupled dynamics of the plasma and the electromagnetic field was derived. Building on this framework, we focus on a black hole spacetime with a topological defect and examine the quasibound states arising for various plasma configurations.

Consider a two-component plasma composed of electrons and ions. We denote the electron number density by $n$ and their four-velocity by $u^\mu$, while $J^\mu$ represents the four-current density associated with the ions. The evolution of the plasma is governed by the following system of differential equations for these quantities \cite{BreuerEhlers1981} as:
\begin{align}
\nabla_\nu F^{\mu\nu} = e n u^\mu + J^\mu,& \label{eq:4} \\
u^\mu \nabla_\mu u^\nu = \frac{e}{m_e} F^\nu\!_\mu u^\mu,& \label{eq:5} \\
u^\mu u_\mu = -1,& \label{eq:6} \\
\nabla_\mu (n u^\mu) = 0&. \label{eq:7}
\end{align}

We study the propagation of perturbations through the plasma by introducing small fluctuations
$\tilde{n}$, $\tilde{u}^{\mu}$, and $\tilde{F}_{\mu\nu}$, for example by decomposing the electromagnetic
field tensor as $F_{\mu\nu} = F^{\text{background}}_{\mu\nu} + \tilde{F}_{\mu\nu},$ and similarly for the other quantities. In this analysis, we neglect second-order perturbations
of both the plasma and the electromagnetic field, as well as perturbations of the background
metric $g_{\mu\nu}$, since the gravitational backreaction of these fields is assumed to be small.
Furthermore, perturbations of the ions are ignored, as they are suppressed relative to electron
perturbations by a factor proportional to $m_e/m_{\text{ion}} \ll 1$. In the case of a hot plasma,
an additional term associated with the fluid pressure would appear in the momentum Eq.~(\ref{eq:5}).

The presence of a plasma implies the existence of a
preferred rest frame. Locally, the plasma defines surfaces of simultaneity for the observer, whose effective metric tensor is
\begin{align}
h_{\mu\nu}=g_{\mu\nu}+u_\mu u_\nu\,.
\end{align}
The tensor $h^{\mu\nu}$ performs an orthogonal projection onto the tangent plane of the rest frame for the (electron) plasma. As a result, the kinematics of the electron fluid are captured by two tensors: the rotation rate (vorticity) $\omega^{\mu\nu} = -\omega^{\nu\mu}$, and the deformation rate $\theta^{\mu\nu} = \theta^{\nu\mu}$. This stems from the standard kinematic decomposition~\cite{Ellis:1971pg}.
\begin{align}
\nabla_\mu u_\nu = \nabla_{(\mu} u_{\nu)} + \nabla_{[\mu} u_{\nu]}=\omega_{\nu\mu}+\theta_{\nu\mu}-u_\mu u^\alpha\nabla_\alpha u_\nu\,,
\end{align}
from which we get
\begin{align}
    \omega_{\mu\nu}&=\frac{1}{2}(v_{\mu\nu}-v_{\nu\mu})\,,\label{eq25}\\
    \theta_{\mu\nu}&=\frac{1}{2}(v_{\mu\nu}+v_{\nu\mu})\,,\label{eq26}
\end{align}
where we have introduced the tensor $v^{\mu\nu} \equiv h^{\mu\alpha} h^{\nu\beta} u_{\alpha;\beta}$. Furthermore, we define the plasma frequency as $\omega_{\rm pl} \equiv (ne^2/m_e)^{1/2}$, the electric component of the electromagnetic field as $E^\mu \equiv F^\mu\!_{\nu} u^\nu$, the magnetic component as $B_{\mu\nu} \equiv h_\mu\!^{\alpha} h_\nu\!^{\beta} F_{\alpha\beta}$, and the Larmor tensor as $\omega_L\!^{\mu\nu} \equiv -\frac{e}{m_e} B^{\mu\nu}$. On the basis of these definitions, by differentiating Maxwell’s equation (\ref{eq:4}) and employing the momentum equation (\ref{eq:5}), Ref.~\cite{BreuerEhlers1981} derived the perturbation equation for the vector potential ${\tilde{A}}^\mu$ in the Landau gauge, $u_\mu \tilde{A}^\mu = 0$, which incorporates both the effects of the gravitational field and those of the moving plasma:
\begin{align}\nonumber
\Bigl(
&h^{\alpha\beta} u^\nu \nabla_\nu (\nabla^\mu\!_\beta - \delta^\mu\!_\beta \nabla^\gamma\!_\gamma)
+ \omega^{\alpha\beta} + \omega_{\mathrm{L}}^{\alpha\beta} + \theta^{\alpha\beta}\\\nonumber&
+ \theta h^{\alpha\beta}
+ \frac{e}{m_e} E^\alpha u^\beta \times (\nabla^\mu\!_\beta - \delta^\mu\!_\beta \nabla^\gamma\!_\gamma)\\&
+ \omega_{\mathrm{pl}}^2 h^{\alpha\mu} u^\gamma \nabla_\gamma
+ \omega_{\mathrm{pl}}^2 (\theta^{\alpha\mu} - \omega^{\alpha\mu})
\Bigr) \tilde{A}_\mu = 0\,, \label{27}
\end{align}
where $\nabla_{\mu\nu}\equiv \nabla_\mu\nabla_\nu$ and $\theta=\theta^\mu\!_{\mu}$. The above relation serves as the point of departure for a systematic and rigorous investigation of the linearized photon dynamics in a cold plasma embedded in a curved spacetime background.

\subsection{EM Waves in Plasma: Dynamical Equations in Black Hole Spacetime with a Topological Defect}

 \begin{figure*}
   \centering
   {\includegraphics[width=0.4\textwidth]{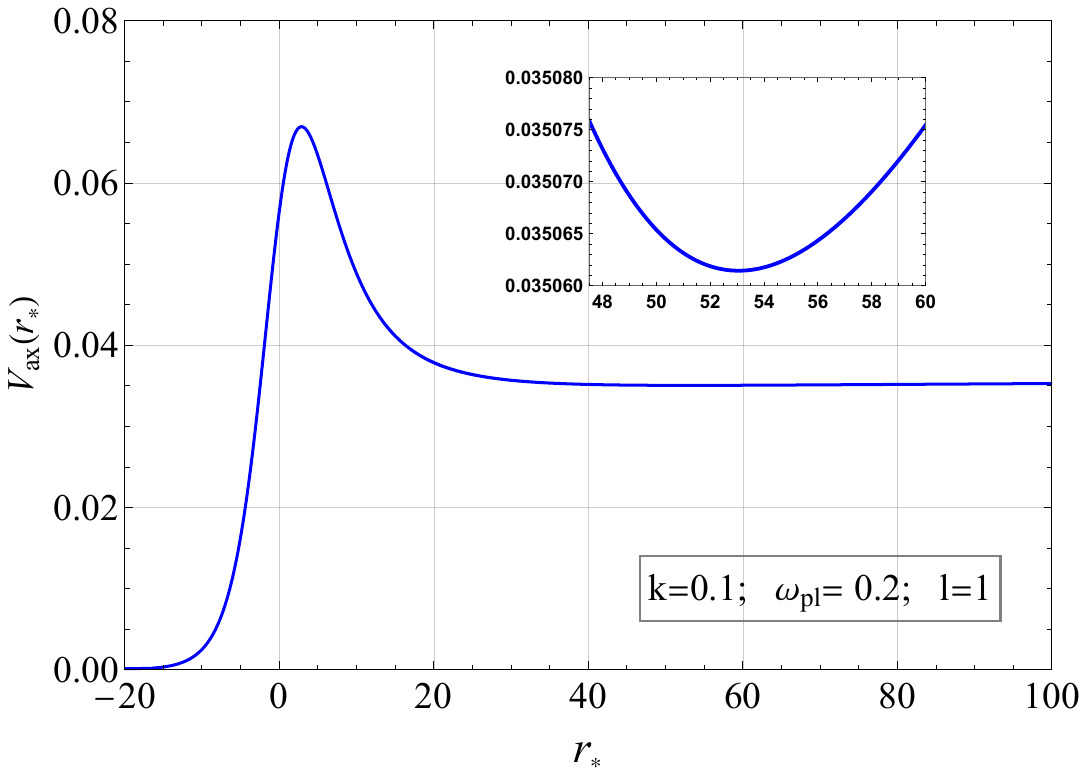}}\hspace{10mm}
   {\includegraphics[width=0.4\textwidth]{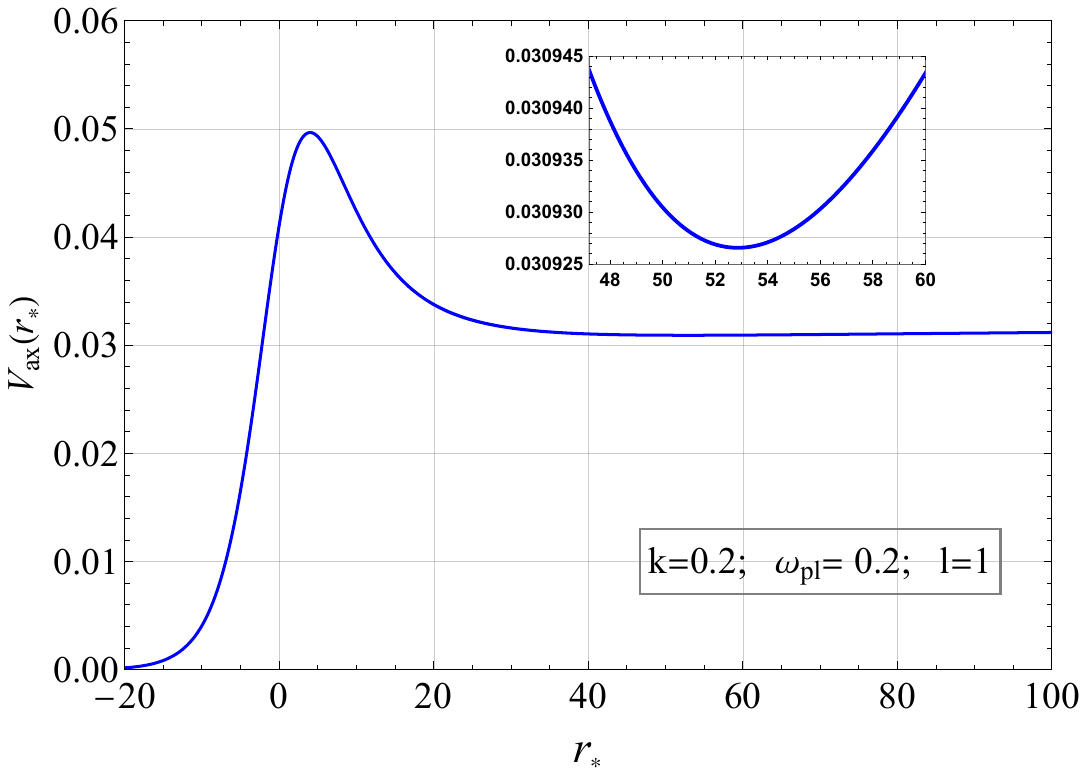}}
   {\includegraphics[width=0.4\textwidth]{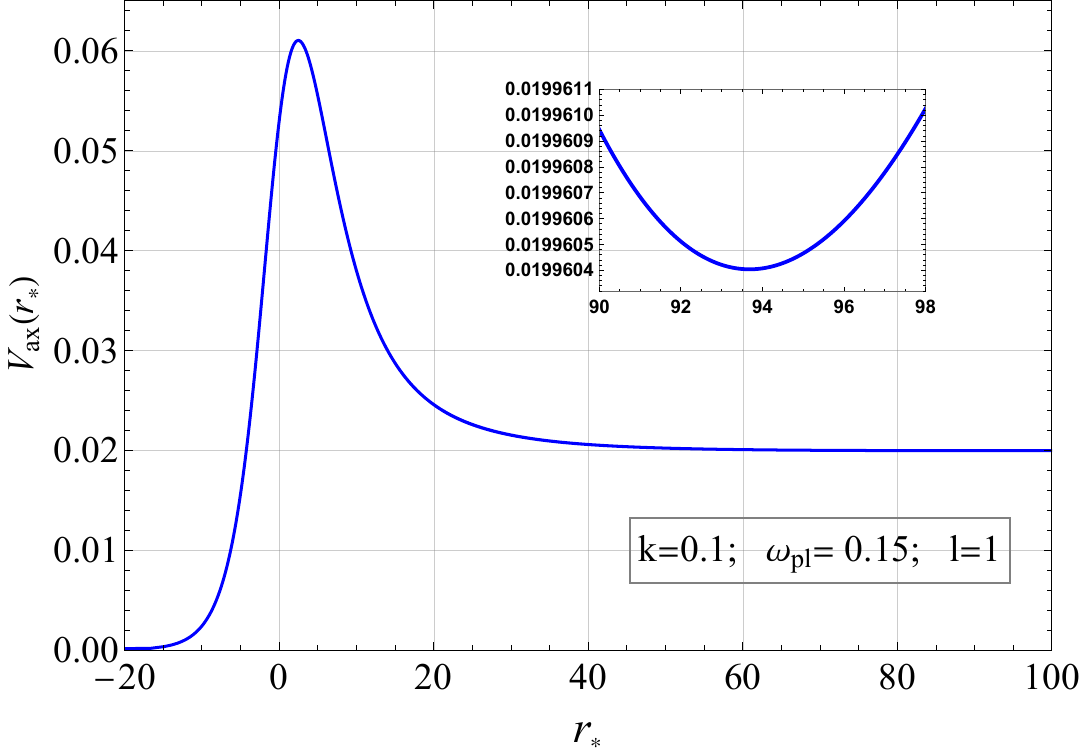}}\hspace{10mm}
   {\includegraphics[width=0.4\textwidth]{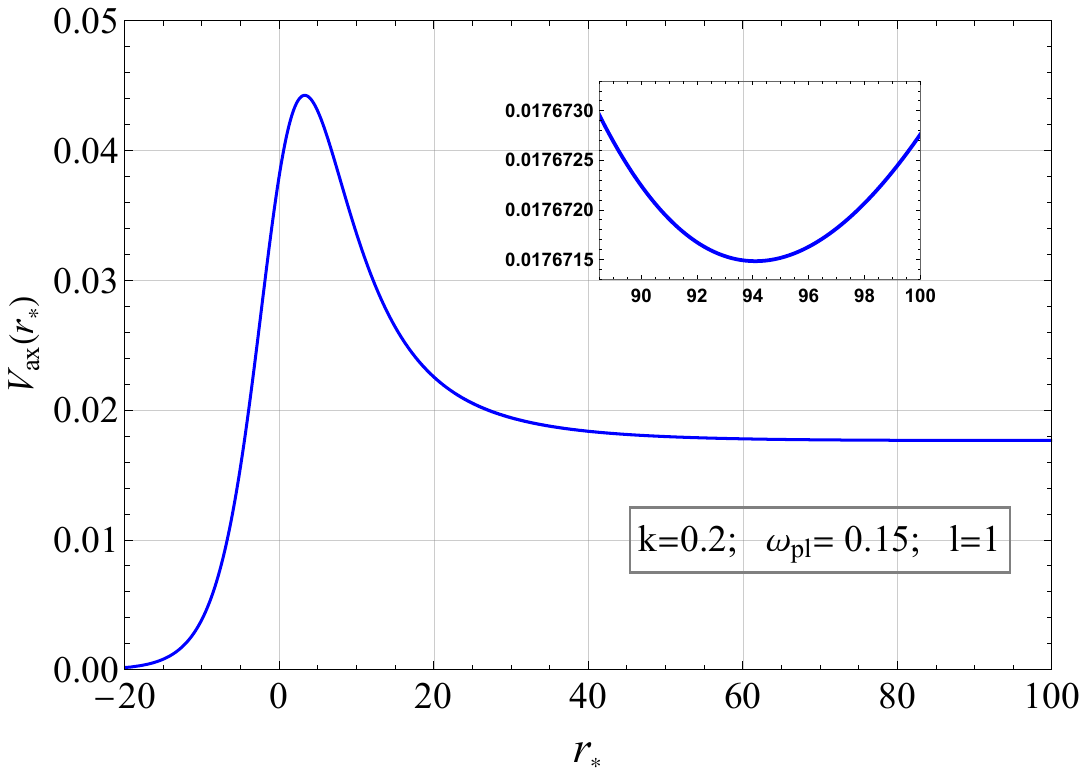}}
   \caption{Plot of the effective potential $V_{ax}$ in the axial sector for a black hole with a topological defect (with $M=1$) as a function of the tortoise coordinate $r_\star$ in a homogeneous plasma medium.} 
   \label{fig6}
\end{figure*}
\begin{figure}
    \centering
    \includegraphics[width=0.85\linewidth]{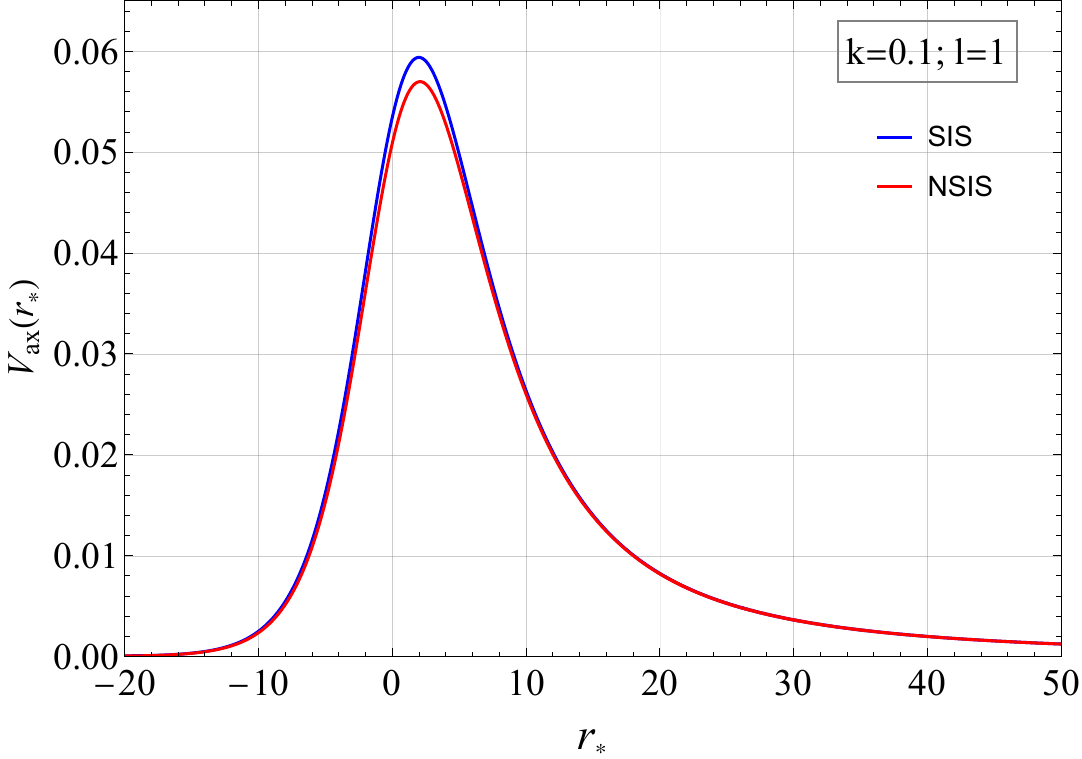}
    \caption{Effective axial potential $ V_{\text{ax}}(r_*) $ for a black hole with a topological defect $( M = 1 )$ in an inhomogeneous plasma medium. Comparison of the SIS model (blue curve) and the NSIS model (red curve) with parameters $K_e\xi=0.2$ and $r_c=3$.
}
    \label{fig7}
\end{figure}
\begin{figure*}
   \centering
   {\includegraphics[width=0.4\textwidth]{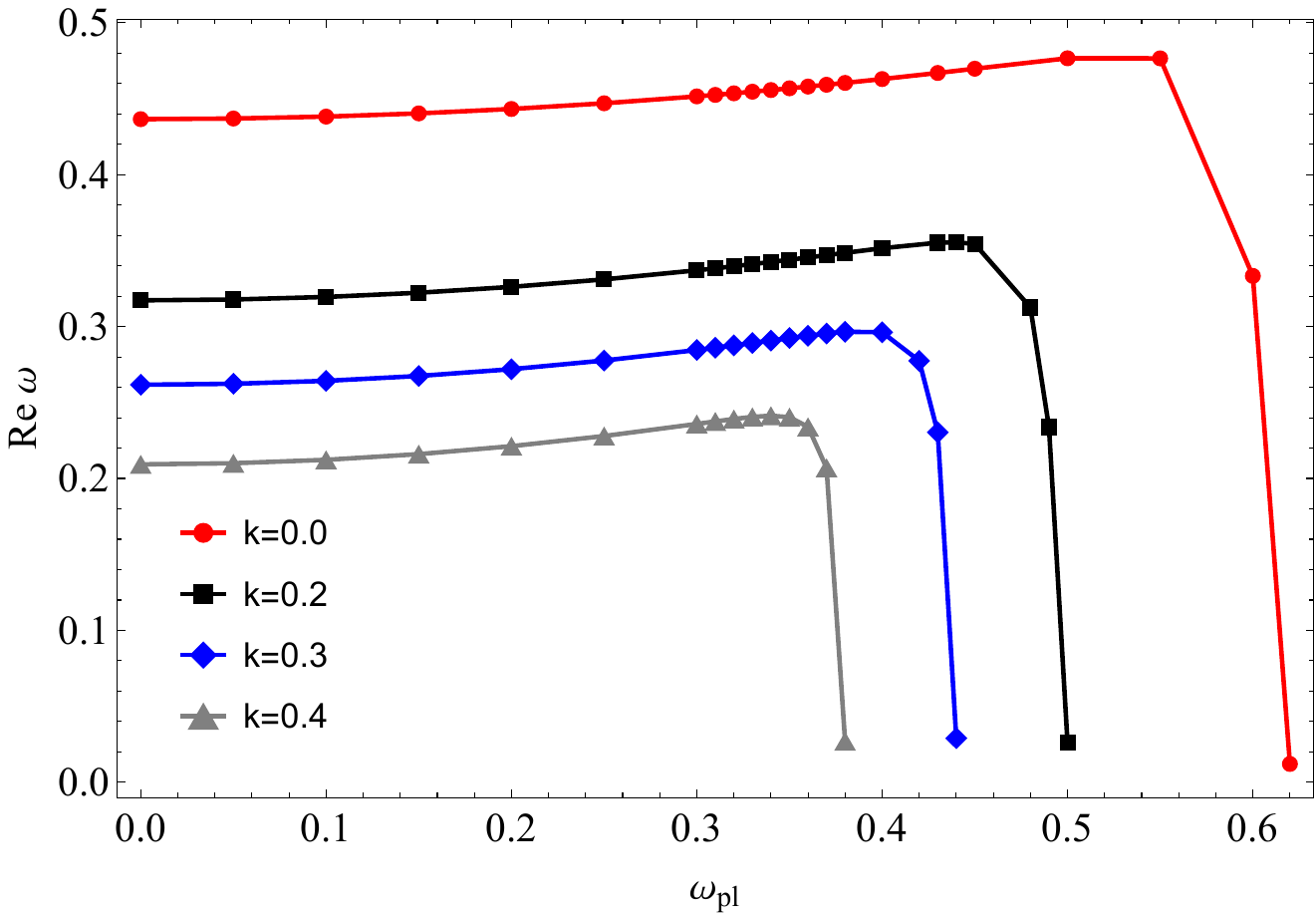}}\hspace{10mm}
   {\includegraphics[width=0.4\textwidth]{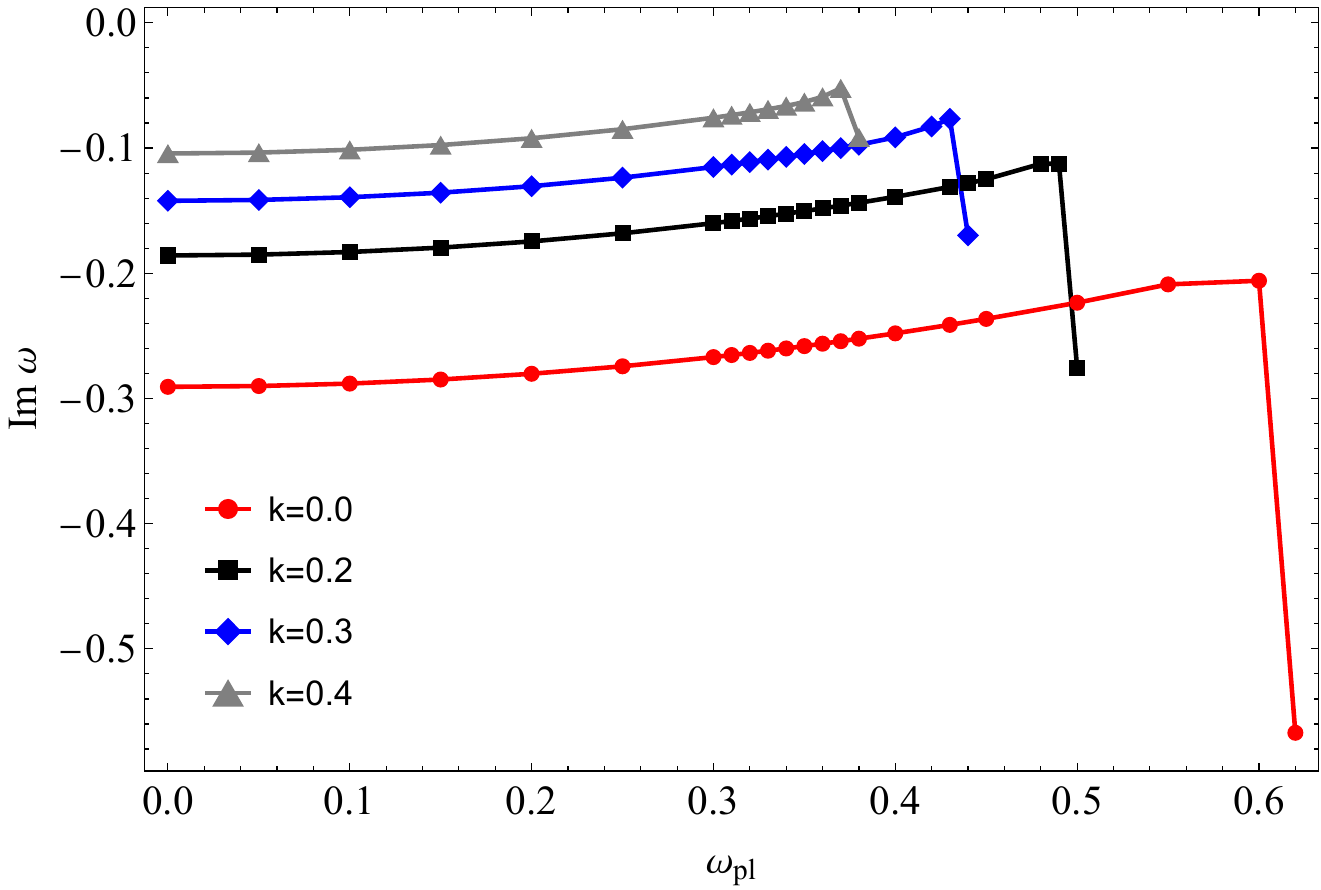}}
\caption{The real and imaginary parts of the frequencies of the fundamental quasi-bound states of electromagnetic perturbations in the axial sector in a homogeneous plasma for $l=2$ and $n=1$ around a black hole with a topological defect. Left panel:  Dependence of $\mathrm{Re}\,\omega$ on the plasma frequency $\omega_{\rm pl}$. Right panel: Dependence of $\mathrm{Im}\,\omega$ on $\omega_{\rm pl}$ for different values of the topological parameter $k$.}
   \label{fig8}
\end{figure*}
%\begin{figure*}
%   \centering
   %{\includegraphics[width=0.4\textwidth]{SINSI1.pdf}}\hspace{10mm}
   %{\includegraphics[width=0.4\textwidth]{SINSI2.pdf}}
  % \caption{\justifying The real and imaginary parts of the frequencies of the fundamental quasi-bound states of electromagnetic perturbations in the axial sector in an inhomogeneous plasma (in the SIS and NSIS models) for $l=2$, $n=1$, and $r_c=3$ around a black hole with a topological defect. The left panel displays the dependence of $\mathrm{Re}\,\omega$  on the plasma frequency $\omega_{\rm pl}$, while the right panel shows the dependence of $\mathrm{Im}\,\omega$ on $\omega_{\rm pl}$ for different values of the topological parameter $k$.
% }
  % \label{s9}
%\end{figure*}
In this section, we derive the dynamical equations governing the propagation of electromagnetic
waves in a plasma surrounding a black hole spacetime with a topological defect.  In
this case both the vorticity and the deformation tensors are zero, as can be easily checked from Eqs. (\ref{eq25} )and (\ref{eq26}). The four velocity of a static plasma is $u^\alpha=(u^0, \vec{0})$, with $u^0=f^{-1/2}$ satisfying the normalization condition $u_\mu u^\mu=-1$. From Eq. (5), the electric field has then only one nonvanishing radial component $E^\alpha=(0,m_e/e \Gamma^r_{00}(u^0)^2,0,0),$ where $\Gamma^{\mu}_{\alpha\beta}$
are the standard Christoffel’s symbols. We assume an unmagnetized
plasma $B_{\mu\nu=0}$ (and therefore also $\omega_{L}\!^{\mu\nu}=0$).

Moreover, in a spherically symmetric spacetime, the angular dependence of the fields was separated from the radial part by means of a multipolar expansion. Following Ref. \cite{Rosa:2011my},  we introduce a basis of four vector spherical harmonics:
\begin{align}
Z^{(1)\ell m} 
&= [1, 0, 0, 0] Y^{\ell m}, 
\label{eq:15} \\
Z^{(1)}_{\mu}{}^{\ell m} 
&= [1, 0, 0, 0] Y^{\ell m}, 
 \\Z^{(2)\ell m} 
&= [0, f^{-1}, 0, 0] Y^{\ell m}, 
\label{eq:16} \\Z^{(3)\ell m}_{\mu} 
&= \frac{r}{\sqrt{\ell(\ell+1)}} 
  [0, 0, \partial_{\theta}, \partial_{\phi}] Y^{\ell m},
 \\Z^{(4)\ell m}_{\mu} 
&= \frac{r}{\sqrt{\ell(\ell+1)}} 
  \Bigl[0, 0, \frac{\partial_{\phi}}{\sin\theta}, 
         -\sin\theta\,\partial_{\theta}\Bigr] Y^{\ell m}.
\end{align}
where $Y^{lm}(\theta,\phi)$ are the standard scalar spherical harmonics. These vector spherical harmonics satisfy the orthogonality condition
\begin{equation}
\int d\Omega \, Z^{(i)\ell m}_{\mu} \hat{\eta}^{\mu\nu} Z^{(i')\ell'm'}_{\nu}
= \delta^{ii'}\delta^{\ell\ell'}\delta^{mm'},
\end{equation}
where $d\Omega=\sin\theta d\theta d\phi$ and $\hat{\eta}^{\mu\nu} = \operatorname{diag}\Bigl[1,\ f^2,\ r^{-2},\ (r\sin\theta)^{-2}\Bigr]$. The perturbation of
the vector potential can be decomposed in this basis as
\begin{align}
\tilde{A}_\mu(r,t,\theta,\phi) = -\frac{1}{r}
\sum_{i=1}^{4} \sum_{\ell=0}^\infty \sum_{m=-\ell}^{\ell}
c_i \, u_{(i)}^{\ell m}(t,r) \, Z^{(i)\,\ell m}_{\mu}(\theta,\phi)\,,
\label{eq:A-decomp}
\end{align}
where $c_1=c_2=1$ and $c_3=c_4=1/\sqrt{l(l+1)}$. Using this decomposition and frequency-domain representation $u^{lm}_{(i)}(t,r)=u^{lm}_{(i)}(r)e^{-i\omega t}$, the field equations become 
\begin{align}
&u_{(1)}=0\,,\label{35}\\ &(rf(l+l^2+r^2\omega^2_{\rm pl})-r^3\omega^2)u_{(2)}-r^2f^2u'_{(3)}=0\,, \label{36} \\  &\nonumber l(l+1)rfu_{(2)}+r^3(\omega^2-f\omega^2_{\rm pl})u_{(3)}-l(1+l)r^2u'_{(2)}\\&+2Mrfu'_{3}+r^3f^2u''_{(3)}=0\,,\label{37}\\&(rf(l+l^2+r^2\omega^2_{\rm pl})+r^3\omega^3)u_{(4)}-2Mrfu'_{(4)}-r^3f^2u''_{(4)}=0\,.\label{38}
\end{align}
Here $u'_{(i)} = \partial_r u'_{(i)}$; the superscript $l$ was omitted, and the plasma frequency was assumed to depend only on the radial coordinate,
$\omega_{\mathrm{pl}} = \omega_{\mathrm{pl}}(r)$. Due to the spherical symmetry of the background spacetime, the equations were independent of the angular number $m$. It is worth noting that, although Eq.~(\ref{27}) contains
third-order derivatives, the resulting system of equations in the frequency
domain reduces to a set of second-order differential equations.

From Eqs.~(\ref{35})-(\ref{38}), it is immediately evident that the polar (even-parity)
sector, described by the functions $u^{(1)}$, $u^{(2)}$, and $u^{(3)}$, is
completely decoupled from the axial (odd-parity) sector, described by the
function $u^{(4)}$. This structure closely resembles the case of a massive
vector field discussed in Ref.~\cite{BreuerEhlers1981} and arises as a direct consequence of the
spherical symmetry of the background. Consequently, the two sectors can be
analyzed independently, as discussed below.
\begin{widetext}
    
\begin{table}
\centering
\caption{The real and imaginary parts of the quasi-bound states of the axial sector of the electromagnetic field in the background of a black hole with a topological defect in a uniform plasma medium. The upper tables correspond to the black hole with a topological defect, while the lower tables correspond to the Schwarzschild case.}
\label{tab:QNM_plasma}
\renewcommand{\arraystretch}{1.3}
\begin{tabular}{|c|c|c|}
\hline
\multicolumn{3}{|c|}{\textbf{$k=0.1$, $\omega_{\rm pl}=0.1$}} \\
\hline
$(l,n)$ & 3rd order WKB & 6th order WKB \\
\hline
$(0,0)$ & $ - $ & $ - $ \\
$(1,0)$ & $0.2184 - 0.0718i$ & $0.2200 - 0.0716i$ \\
$(1,1)$ & $0.1854 - 0.2349i$ & $0.1878 - 0.2333i$ \\
$(2,0)$ & $0.3952 - 0.0759i$ & $0.3955 - 0.0758i$ \\
$(2,1)$ & $0.3771 - 0.2328i$ & $0.3776 - 0.2327i$ \\
$(2,2)$ & $0.3490 - 0.3985i$ & $0.3479 - 0.4025i$ \\
\hline
\end{tabular}
\begin{tabular}{|c|c|c|}
\hline
\multicolumn{3}{|c|}{\textbf{$k=0.15$, $\omega_{\rm pl}=0.15$}} \\
\hline
$(l,n)$ & 3rd order WKB & 6th order WKB \\
\hline
$(0,0)$ & $ - $ & $ - $ \\
$(1,0)$ & $0.2101 -0.0589i$ & $0.2115 -0.0587i$ \\
$(1,1)$ & $0.1723 -0.2041i$ & $0.1758 -0.2011i$ \\
$(2,0)$ & $0.3684 -0.0660i$ & $0.3687 -0.0660i$ \\
$(2,1)$ & $0.3503 -0.2038i$ & $0.3508 -0.2037i$ \\
$(2,2)$ & $0.3232 -0.3517i$ & $0.3224 -0.3547i$ \\
\hline
\end{tabular}
\begin{tabular}{|c|c|c|}
\hline
\multicolumn{3}{|c|}{\textbf{$k=0.2$, $\omega_{\rm pl}=0.2$}} \\
\hline
$(l,n)$ & 3rd order WKB & 6th order WKB \\
\hline
$(0,0)$ & $ - $ & $ - $ \\
$(1,0)$ & $0.2062 - 0.0442i$ & $0.2073 - 0.0441i$ \\
$(1,1)$ & $0.1582 - 0.1748i$ & $0.1691 - 0.1672i$ \\
$(2,0)$ & $0.3448 - 0.0560i$ & $0.3450 - 0.0560i$ \\
$(2,1)$ & $0.3256 - 0.1748i$ & $0.3261 - 0.1746i$ \\
$(2,2)$ & $0.2980 - 0.3063i$ & $0.2979 - 0.3079i$ \\
\hline
\end{tabular}

\begin{tabular}{|c|c|c|}
\hline
\multicolumn{3}{|c|}{Shw, $\omega_{\rm pl}=0.1$} \\
\hline
$(l,n)$ & 3rd order WKB & 6th order WKB \\
\hline
$(0,0)$ & $ - $ & $ - $ \\
$(1,0)$ & $0.2515 -0.0893i$ & $0.2539 -0.0889i$ \\
$(1,1)$ & $0.2109 -0.2921i$ & $0.2142 -0.2899i$ \\
$(2,0)$ & $0.4606 -0.0938i$ & $0.4610 -0.0938i$ \\
$(2,1)$ & $0.4375 -0.2884i$ & $0.4382 -0.2881i$ \\
$(2,2)$ & $0.4023 -0.4937i$ & $0.4010 -0.4995i$ \\
\hline
\end{tabular}
\begin{tabular}{|c|c|c|}
\hline
\multicolumn{3}{|c|}{Shw, $\omega_{\rm pl}=0.15$} \\
\hline
$(l,n)$ & 3rd order WKB & 6th order WKB \\
\hline
$(0,0)$ & $ - $ & $ - $ \\
$(1,0)$ & $0.2586 -0.0844i$ & $0.2610 -0.0841i$ \\
$(1,1)$ & $0.2103-0.2882i$ & $0.2145 -0.2845i$ \\
$(2,0)$ & $0.4649 -0.0923i$ & $0.4654 -0.0923i$ \\
$(2,1)$ & $0.4396 -0.2851i$ & $0.4403 -0.2849i$ \\
$(2,2)$ & $0.4022 -0.4912i$ & $0.4011 -0.4967i$ \\
\hline
\end{tabular}
\begin{tabular}{|c|c|c|}
\hline
\multicolumn{3}{|c|}{Shw, $\omega_{\rm pl}=0.2$}  \\
\hline
$(l,n)$ & 3rd order WKB & 6th order WKB \\
\hline
$(0,0)$ & $ - $ & $ - $ \\
$(1,0)$ & $0.2687 -0.0775i$ & $0.2711 -0.0772i$ \\
$(1,1)$ & $0.2090 -0.2846i$ & $0.2175 -0.2761i$ \\
$(2,0)$ & $0.4710 -0.0901i$ & $0.4714 -0.0901i$ \\
$(2,1)$ & $0.4424 -0.2806i$ & $0.4433 -0.2803i$ \\
$(2,2)$ & $0.4020 -0.4879i$ & $0.4013 -0.4927i$ \\
\hline
\end{tabular}

\end{table}

\end{widetext}
\subsection{ Quasi-bound states:  Axial sector}
In the axial sector, we have a single equation for $u_{(4)}$
to solve. Thus, the axial sector can be
easily reduced to a Schr\"odinger-like
 equation analogous to the massive vector case
\begin{equation}
\mathcal{D} u_{(4)}(r) = 0, 
\end{equation}
with
\[
\mathcal{D} \equiv \frac{d^2}{dr^2} + \omega^2 - f\left( \frac{l(l+1)}{r^2} + \omega_{\rm pl}^2 \right),
\]
and in terms of the tortoise coordinate defined by $\frac{dr_*}{dr} = f^{-1}.$
In this case the plasma frequency plays indeed the role of an effective mass for the component $u_{(4)}$.

The effective potential for the axial part takes the following form for a black hole with a topological defect.
\begin{align}
V_{\ell m}^{\rm ax}(r) = \bigg(1-k-\frac{2M}{r}\bigg) \bigg( \frac{\ell(\ell+1)}{r^2} +\omega_{\rm pl}^2 \bigg)\,.\label{eq47}
\end{align}
Intriguingly, the axial potential does not depend on $m$, and in the limit $\omega_{\rm pl} \rightarrow 0$, it reduces to the effective potential experienced by a photon in the background of a black hole with a topological defect.

Using the semiclassical WKB method, we will determine the quasi-bound states associated with the axial sector of electromagnetic waves. We start by assuming a homogeneous plasma profile, for which the effective potential of a black hole with a topological defect is given as defined in Eq. (\ref{eq47}).

For the existence of a quasi-bound state around black holes with a topological defect, the effective potential must have two extrema one maximum and one minimum, where the maximum must be located closer to the compact object than the minimum (see Fig. \ref{fig6} for an example). From Fig.~\ref{fig6} it is not difficult to see that the position and the depth of the potential well depend significantly on the plasma frequency $\omega_{\rm pl}$ and on the topological defect parameter $k$. As the plasma frequency decreases, the height of the potential barrier decreases, while the potential well becomes deeper for fixed values of $k$ and the multipole number $l$ (left panels). In addition, with decreasing plasma frequency the position of the potential well shifts toward larger values of the tortoise coordinate $r_{*}$, i.e., it moves farther away from the potential barrier.

On the other hand, increasing the topological defect parameter $k$ for fixed values of $\omega_{\rm pl}$ and the multipole number $l$ (upper panels) leads to a decrease in the height of the potential barrier and modifies its shape. At the same time, the position of the potential well remains almost unchanged relative to the potential barrier. 

Taking the derivative of the effective axial potential $V_{\ell m}^{\rm ax}$, one can show that the above conditions are satisfied for $M \omega_{\mathrm{pl }}\leq (1-k)\sqrt{l(l+1)/12}$ (for a derivation, see Appendix \ref{AppC}). Thus, there exists an upper bound for the
plasma frequency.

Fig. \ref{fig7} shows a comparison of the effective axial potential of electromagnetic perturbations for the SIS and NSIS plasma profiles. The plot shows that the differences between the two models are small however, the SIS model leads to a slightly higher potential maximum compared to the NSIS model. This behavior is due to the steeper plasma density profile in the SIS distribution, which enhances the effective refractive contribution of the plasma near the black hole. As a result, electromagnetic perturbations experience a somewhat stronger scattering potential barrier in the SIS case compared to the NSIS model.

In the SIS model, the plasma frequency scales as $\omega_{\rm pl}^2 \propto \frac{1}{r^2}$. This dependence is functionally identical to the centrifugal barrier $l(l+1)/r^2$. Therefore, the plasma contribution does not change the qualitative shape of the effective potential, but only leads to a renormalization of the angular momentum,
$l(l+1) \to l(l+1) + \alpha$,
where the parameter $\alpha=K_e\xi$ is determined by the properties of the plasma medium. As a result, the effective potential retains the same structure as in the vacuum case it possesses a single maximum near the photon sphere, after which it monotonically decreases and approaches zero at spatial infinity.

In the NSIS model, quasi-bound states also do not arise. Although the plasma density profile in this model differs near the center, at large distances \(r \gg r_c\) the asymptotic behavior of the plasma frequency coincides with that in the SIS model and also scales as
$\omega_{\rm pl}^2 \propto \frac{1}{r^2}$. As a consequence, the effective potential at large distances has the same functional form. The potential possesses only a single maximum near the photon sphere and does not form a potential well. Thus, the trapping region necessary for the existence of quasi-bound states is absent, and the conditions for their formation are not satisfied in either plasma model.

Fig. \ref{fig8} shows the real $\mathrm{Re}\,\omega$ (left panel) and imaginary $\mathrm{Im}\,\omega$ (right panel) parts of the frequencies of the fundamental quasi-bound states of electromagnetic perturbations in the axial sector near a black hole with a topological defect in a homogeneous plasma. Different curves correspond to different values of the topological parameter $k$.

The left panel demonstrates that the real part of the frequency, $\mathrm{Re}\,\omega$, generally increases with the plasma frequency $\omega_{\rm pl}$, but when $\omega_{\rm pl}$ exceeds a critical value, a sharp cutoff is observed, indicating the disappearance of quasi-bound states. This is consistent with the conclusion from the equation provided: both roots exist only if the condition  $M \omega_{\mathrm{pl }}\leq (1-k)\sqrt{l(l+1)/12}$ is satisfied, which was derived in Appendix \ref{AppC}. Therefore, the critical plasma frequency at which quasi-bound states cease to exist depends directly on the topological parameter $k$ and the multipole number $l$.

The right panel shows that as  $\omega_{\rm pl}$ increases, the magnitude $|\mathrm{Im}\,\omega|$ slightly decreases before the sharp cutoff, reflecting a weakening of the damping of the quasi-bound states as the plasma frequency approaches the threshold. Higher values of $k$ generally correspond to smaller $|\mathrm{Im}\,\omega|$, indicating a longer-lived perturbation.

Table \ref{tab:QNM_plasma} presented the real and imaginary parts of the quasi-bound state frequencies of the axial sector of the electromagnetic field in the background of a black hole with a topological defect (upper tables) and in the Schwarzschild background (lower tables) in a homogeneous plasma medium. The calculations were performed using the third- and sixth-order WKB methods for different values of the multipole number $l$  and the overtone number  $n$.

\subsection{Polar sector}
The polar equations governing $u_{(2)}$ and $u_{(3)}$ can be combined into a single Schrödinger-like equation
\begin{equation}
\frac{d^2\psi}{dr_*^2} - V(r)\psi = 0\,, \label{48}
\end{equation}
where the complicated form of the effective potential $V(r)$ is given Appendix \ref{Ap2}. The reduction of the polar sector to a single second-order differential equation indicates that only one dynamical degree of freedom is present. The effective potential 
$V$ depends on the plasma frequency $\omega_{\rm pl}$ as well as on its radial derivatives. 

In this article, we focused on the WKB method however, Eq. (\ref{48}) could not be solved due to the complicated dependence of the effective potential on $\omega$. To solve Eq. (\ref{48}), other methods were required, which had already been studied in papers \cite{Cannizzaro:2020uap,Chowdhury:2024auw}.

\section{Conclusions}
\label{sec:conclusion}

In this study, we examine the impact of a plasma environment on the optical and perturbative characteristics of a black hole endowed with a topological defect. Our investigation encompasses three closely interrelated components: the correspondence between the black hole shadow and QNMs in the eikonal limit; the QNM spectrum of a massive scalar field propagating in both homogeneous and inhomogeneous plasma distributions; and the quasi-bound configurations of electromagnetic perturbations evolving within a cold plasma medium.

First, we studied the shadow--QNM correspondence in the presence of plasma and showed that the refractive index modifies the relation between the real part of the eikonal QNM frequency and the shadow radius. We also analyzed the Lyapunov exponent associated with unstable photon-sphere orbits and found that it depends only weakly on the plasma frequency, while it decreases systematically as the topological-defect parameter $k$ increases. This indicates that the topological defect reduces the instability timescale of circular null geodesics and therefore affects the characteristic oscillation scale encoded in the eikonal spectrum.

Second, we examined massive scalar-field perturbations in homogeneous plasma and in two inhomogeneous plasma models, namely the SIS and NSIS profiles. By deriving the corresponding effective potential and applying the third- and sixth-order WKB methods, we found that both the plasma distribution and the topological-defect parameter modify the scalar QNM spectrum. In particular, increasing $k$ leads to a monotonic decrease in the real part of the frequencies, while the plasma changes both the oscillation frequency and damping rate through its contribution to the effective potential. A comparison among plasma models shows that the SIS and NSIS cases produce only small quantitative differences, with the NSIS profile generally yielding slightly higher oscillation frequencies than the SIS and homogeneous cases.

Third, we derived the dynamical equations governing electromagnetic perturbations in a cold, unmagnetized plasma and showed that the axial and polar sectors decouple. In the axial sector, the plasma frequency plays the role of an effective mass term, which makes possible the existence of quasi-bound states in the homogeneous plasma case. We showed that such states exist only below a critical plasma-frequency threshold determined by the topological-defect parameter and the multipole number. As the plasma frequency approaches this critical value, the quasi-bound spectrum terminates. In contrast, for the SIS and NSIS plasma profiles, the effective axial potential does not develop the barrier--well structure required for trapping, and quasi-bound states are therefore absent.

Overall, the findings presented here indicate that plasma contributions and topological defects generate correlated yet distinguishable imprints on shadow observables, scalar QNMs spectra, and electromagnetically trapped configurations. These characteristic signatures may offer a viable means of probing both nonvacuum astrophysical environments and nontrivial spacetime structures in the vicinity of compact objects. A natural extension of this work would be to analyze the gravitational perturbation sector directly and to assess the robustness of the scalar QNMs results by employing complementary numerical techniques that go beyond the WKB approximation, particularly in the low-multipole regime.

\section{ACKNOWLEDGMENTS}
This research is partly supported by Research Grant F-FA-2021-510 from the Ministry of Higher Education, Science and Innovation of the Republic of Uzbekistan.

\appendix
\section{Photon-Sphere Radius and Lyapunov Exponent}\label{Ap1}
\begin{align}
\frac{r_{ps}}{M}&=\frac{ -\frac{3}{2} + 2 (1 - k)\eta - \sqrt{ \frac{9}{4} - 2\eta (1 - k)  } }{ (1 - k) \big( (1 - k) \eta - 1 \big) }\,,
\end{align}
\begin{align}
\lambda=\frac{2 (1 - k)^2 \big(1 - (1-k)\eta\big) \sqrt{ \sqrt{9 - 8 (1-k)\eta} - 1 } }{ M \left( 3 + \sqrt{9 - 8 (1-k)\eta} - 4\eta (1-k)\right)^{3/2} }\,.
\end{align}
\section{}\label{Ap2}
\subsection{Effective potential for the polar sector in the homogeneous plasma case}
\begin{widetext}
\begin{align}\nonumber
V_{\rm pl}=-A^{-1}\bigg[&4 r^{12} \omega^{10}
+ 4 l (1 + l) (2 M + (-1 + k) r)^6 \omega_{\text{pl}}^4 \bigl( l + l^2 + 4 r^2 \omega_{\text{pl}}^2 \bigr) 
- 4 r^9 (-2 M + r - k r) \omega^8 \bigl( 3 l (1 + l) + 5 r^2 \omega_{\text{pl}}^2 \bigr) \\ \nonumber
&+ 4 r (2 M + (-1 + k) r)^5 \omega_{\text{pl}}^4 \Bigl(
  l^2 (1 + l)^2 (1 + l + l^2) + l (1 + l) (1 + 3 l (1 + l)) r^2 \omega_{\text{pl}}^2 + 3 l (1 + l) r^4 \omega_{\text{pl}}^4 + r^6 \omega_{\text{pl}}^6
\Bigr) \\ \nonumber
&+ r^4 (2 M + (-1 + k) r) \omega^6 \Bigl(
  2 l (1 + l) r^3  + 30 l (1 + l) r (2 M + (-1 + k) r)^2 + (2 M + (-1 + k) r) \bigl(
    4 l (1 + l) (5  \\ \nonumber
    &+ 3 l (1 + l)) r^2 + 48 l (1 + l) r^4 \omega_{\text{pl}}^2 + 40 r^6 \omega_{\text{pl}}^4
  \bigr)
\Bigr) + 2 r (2 M + (-1 + k) r)^3 \omega^2 \omega_{\text{pl}}^2 \Bigl(
  -l (1 + l) r^2 (l + l^2 + r^2 \omega_{\text{pl}}^2) \\ \nonumber
  & + l (1 + l) (2 M + (-1 + k) r)^2 (3 l (1 + l) + 29 r^2 \omega_{\text{pl}}^2)  + 2 r (2 M + (-1 + k) r) \Bigl(
    l^2 (1 + l)^2 (1 + 2 l (1 + l)) \\ \nonumber
    & + l (1 + l) (5 + 9 l (1 + l)) r^2 \omega_{\text{pl}}^2 + 12 l (1 + l) r^4 \omega_{\text{pl}}^4 + 5 r^6 \omega_{\text{pl}}^6
  \Bigr)
\Bigr) \\ \nonumber
&- r^2 (2 M + (-1 + k) r)^2 \omega^4 \Bigl(
  l^2 (1 + l)^2 r^2 - 3 l (1 + l) (2 M + (-1 + k) r)^2 (l + l^2 + 24 r^2 \omega_{\text{pl}}^2)\\&- 2 r (2 M + (-1 + k) r) \Bigl(
    l^2 (1 + l)^2 (1 + 2 l (1 + l))  + 18 l (1 + l) (1 + l + l^2) r^2 \omega_{\text{pl}}^2  + 36 l (1 + l) r^4 \omega_{\text{pl}}^4 + 20 r^6 \omega_{\text{pl}}^6
  \Bigr)
\Bigr)\bigg]
\end{align} 
where
\begin{align}
  A=  4 r^6 \bigg( \omega^2 + \dfrac{(2 M + (k-1 ) r) \omega_{\text{pl}}^2}{r} \bigg)
\bigl( l (1 + l) (r-2 M  - k r) - r^3 \omega^2 + r^2 (r-2 M - k r) \omega_{\text{pl}}^2 \bigr)
\end{align}
\end{widetext}

\subsection{Effective potential for the polar sector in the inhomogeneous plasma case
}
In the general case of an inhomogeneous plasma, the effective potential for the polar sector was discussed in Ref. \cite{Chowdhury:2024auw}.

\section{Relative location of maxima and minima for the axial effective potential}\label{AppC}
The effective axial potential $V_{\ell m}^{\rm ax}(r)$ is given by Eq. (\ref{eq47}), which admits both a minimum and maximum. The maximum of the potential corresponds to the light ring, whereas the minimum arises specifically from the effective mass associated with the axial sector of the electromagnetic field. For the existence of quasi-bound orbits, the minimum must be located at larger radial coordinate than maximum, which imposes a condition on the parameters of the problem. By setting $V_{\ell m}^{\rm ax'}(r)=0$, we obtain the following algebraic equation 
\begin{equation}
r^2 - \frac{l(l+1)(1-k)}{M \omega_{\mathrm{pl}}^2}\, r + \frac{3l(l+1)}{\omega_{\mathrm{pl}}^2} = 0\,,
\end{equation}
whose solutions depict the location of the extrema of the potential $V_{\ell m}^{\rm ax}$. As evident from the above equation, both the roots will exist if the following condition holds
\begin{equation}
M \omega_{\mathrm{pl }}\leq (1-k)\sqrt{\frac{l(l+1)}{12}}\,.
\end{equation}
The above inequality suggests that for $l=1$, $k=0.1$ for the existence of double root, we must have $M\omega_{\mathrm{pl }}\leq (9/10\sqrt{6})$. When this condition is satisfied, the roots of the above algebraic equation read,

\begin{widetext}
\begin{align}
r_{>} &= \frac{1}{2} \left[ \frac{l(l+1)(1-k)}{M \omega_{\mathrm{pl}}^2} 
+ \sqrt{ \frac{l(l+1)}{\omega_{\mathrm{pl}}^2} \Biggl( \frac{l(l+1)(1-k)^2}{M^2 \omega_{\mathrm{pl}}^2} - 12 \Biggr) } \right], \\
r_{<} &= \frac{1}{2} \left[ \frac{l(l+1)(1-k)}{M \omega_{\mathrm{pl}}^2} 
- \sqrt{ \frac{l(l+1)}{\omega_{\mathrm{pl}}^2} \Biggl( \frac{l(l+1)(1-k)^2}{M^2 \omega_{\mathrm{pl}}^2} - 12 \Biggr) } \right].
\end{align}
\end{widetext}
It is easy to verify that $r_{>} > r_{<}$, where $r_{>}$ is the position of the minimum and $r_{<}$ is the position of the maximum. This means that the minimum is always located at a larger distance than the maximum.
Therefore, considering the monopole mode, as long as $(M  \omega_{\rm pl} \leq 9/10\sqrt{6} \simeq 0.367),$ there exists a potential well outside the light ring, leading to the formation of quasi-bound states.

\newpage
\bibliographystyle{apsrev4-1}
\bibliography{cite}
\end{document}